\pdfoutput=1

\documentclass[11pt,twoside,a4paper,cmspaper,final,collab]{cms-tdr}
\def\svnVersion{8076fbc-D}\def\svnDate{2020/02/18}\def\cmsCernNoTag{CERN-EP-2019-170}\def\cmsCernDate{\today}\def\cmsMessage{"Published in the Journal of High Energy Physics as \href{http://dx.doi.org/10.1007/JHEP03(2020)103}{\doi{10.1007/JHEP03(2020)103}.}"}

\begin{document}\cmsNoteHeader{HIG-18-017}

\hyphenation{had-ron-i-za-tion}
\hyphenation{cal-or-i-me-ter}
\hyphenation{de-vices}
\providecommand{\NA}{\ensuremath{\text{---}}}
\providecommand{\CL}{CL\xspace}
\newlength\cmsTabSkip
\setlength\cmsTabSkip{2ex}
\providecommand{\cmsTable}[1]{\resizebox{\textwidth}{!}{#1}}

\newcommand{\Hmt} {\ensuremath{\PH \to \mu \tau}\xspace}
\newcommand{\Het} {\ensuremath{\PH \to \Pe\tau}\xspace}
\newcommand{\Hmue}{\ensuremath{\PH \to \PGm \PGt_{\Pe}}\xspace}
\newcommand{\Hmuhad}{\ensuremath{\PH \to \PGm \PGt_{\PSh}}\xspace}
\newcommand{\Hehad}{\ensuremath{\PH \to \Pe \PGt_{\PSh}}\xspace}
\newcommand{\Hemu}{\ensuremath{\PH \to \Pe \PGt_{\PGm}}\xspace}
\newcommand{\mue}{\ensuremath{\PGm \PGt_{\Pe}}\xspace}
\newcommand{\emu}{\ensuremath{\Pe \PGt_{\PGm}}\xspace}
\newcommand{\muhad}{\ensuremath{\PGm \PGt_{\PSh}}\xspace}
\newcommand{\ehad}{\ensuremath{\Pe \PGt_{\PSh}}\xspace}
\newcommand{\mcol}{\ensuremath{M_{\text{col}}}\xspace}
\newcommand{\mvis}{\ensuremath{M_{\text{vis}}}\xspace}
\newcommand{\wjets}{\ensuremath{\PW\!+\!\text{jets}}\xspace}
\newcommand{\zjets}{\ensuremath{\PZ\!+\!\text{jets}}\xspace}

\cmsNoteHeader{HIG-18-017}
\title{Search for lepton flavour violating decays of a neutral heavy Higgs boson to \texorpdfstring{$\mu\tau$  and  $\Pe\tau$ in proton-proton collisions at $\sqrt{s}=13$\TeV}{mu tau  and   e tau in proton-proton collisions at sqrt(s)=13 TeV}}

\date{\today}

\abstract{
A search for lepton flavour violating decays of a neutral non-standard-model Higgs boson in the $\mu\tau$ and $\Pe\tau$ decay modes is presented. The search is based on proton-proton collisions at a center of mass energy $\sqrt{s}=13$\TeV collected with the CMS detector in 2016, corresponding to an integrated luminosity of $35.9\fbinv$. The $\tau$ leptons are reconstructed in the leptonic and hadronic decay modes. No signal is observed in the mass range 200--900\GeV. At 95\% confidence level, the observed (expected) upper limits on the production cross section multiplied by the branching fraction vary from 51.9 (57.4)\unit{fb} to 1.6 (2.1)\unit{fb} for the $\mu\tau$ and from 94.1 (91.6)\unit{fb} to 2.3 (2.3)\unit{fb} for the $\Pe\tau$ decay modes.
}

\hypersetup{%
pdfauthor={CMS Collaboration},%
pdftitle={Search for lepton flavour violating decays of a neutral heavy Higgs boson to  mu tau  and   e tau in proton-proton collisions at sqrt(s)=13 TeV},%
pdfsubject={CMS},%
pdfkeywords={CMS, physics, Higgs, taus}
}

\maketitle

\section{Introduction}
The discovery of the 125\GeV Higgs boson, $\PH(125)$, at the CERN LHC in
2012~\cite{Aad:2012tfa, Chatrchyan:2012ufa, Chatrchyan:2013lba} was a major breakthrough
in particle physics. A combined study of data from collisions at $\sqrt{s}$ =  7 and 8\TeV
collected by the ATLAS and CMS Collaborations shows the particle to have properties
consistent with the standard model (SM) Higgs
boson~\cite{Englert:1964et,Higgs:1964ia,Higgs:1964pj,Guralnik:1964eu,Higgs:1966ev,Kibble:1967sv}
including the spin, couplings, and charge-parity
assignment~\cite{JHEP2016:45,Sirunyan:2018koj}. Lepton flavour violating (LFV) decays of
the $\PH(125)$ are forbidden in the SM. However, the presence of new physics in the Higgs
sector is not excluded~\cite{Buschmann:2016pb} and there exist many possible extensions of
the SM that allow LFV decays of the $\PH(125)$. These include the two Higgs doublet
model~\cite{PhysRevLett.38.622}, supersymmetric
models~\cite{DiazCruz:1999xe,Han:2000jz,Arganda:2004bz,Arhrib:2012ax,Arana-Catania:2013xma,Arganda:2015uca,Arganda:2015naa},
composite Higgs  models~\cite{Agashe:2009di,Azatov:2009na}, models with flavour
symmetries~\cite{Ishimori:2010au}, Randall--Sundrum
models~\cite{Perez:2008ee,Casagrande:2008hr,Buras:2009ka}, and
others~\cite{Blanke:2008zb,Giudice:2008uua,AguilarSaavedra:2009mx,Albrecht:2009xr,Goudelis:2011un,McKeen:2012av,Pilaftsis199268,PhysRevD.47.1080,Arganda:2014dta}.
A common feature of many of these models is the presence of additional neutral Higgs
bosons ($\PH$ and A) that would also have LFV decays~\cite{PhysRevD.93.055021,Arganda_2019}.

The most recent search for LFV decays of the $\PH(125)$ was performed by the CMS Collaboration  in the $\PGm\PGt$ and $\Pe\PGt$ channels, using proton-proton ($\Pp\Pp$) collision data recorded at a centre-of-mass energy of $\sqrt{s}=13$\TeV, and corresponding to an integrated luminosity of 35.9\fbinv~\cite{HIG-17-001}. The observed\,(expected) upper limits set on the branching fractions were $\mathcal{B}(\PH(125)\to\PGm\PGt)<0.25\,(0.25)$\% and $\mathcal{B}(\PH(125)\to\Pe\PGt)<0.61\,(0.37)$\% at 95\% confidence level (\CL). These constraints were  a significant improvement over the previously set limits by the CMS and ATLAS Collaborations using the 8\TeV $\Pp\Pp$ collision data set, corresponding to an integrated luminosity of 20\fbinv~\cite{Khachatryan:2015kon,HIG-14-040,Aad:2016blu,Aad:2015gha}. Results from the previous CMS $\PH(125) \to \PGm \PGt$ search, performed using 8\TeV $\Pp\Pp$ collision data, were used to set limits on high mass LFV \PH decays in a phenomenological study~\cite{Buschmann:2016pb}. Limits on the product of the production cross section with branching fraction for the $\PH \to \PGm \PGt$ channel were obtained for \PH mass, $m_{\PH}$, less than 300\GeV.

This paper describes the first direct search for LFV \Hmt and \Het decays for an \PH mass in  the range $200<m_{\PH}<900$\GeV. The search is performed in four decay channels, $\PH \to \PGm \tauh$, $\PH \to \PGm \PGt_{\Pe}$, $\PH \to \Pe \tauh$, and $\PH \to \Pe \PGt_{\PGm}$ where $\tauh$, $\PGt_{\Pe}$, and $\PGt_{\PGm}$ correspond to the hadronic, electronic and muonic decay channels of $\tau$ leptons, respectively. The final-state signatures are  very similar to those of the $\PH \to \PGt\PGt$ decays, studied by CMS~\cite{Chatrchyan:2014vua,CMS-PAPERS-HIG-13-004,CMS-PAS-HIG-16-043,mssmHiggs2018} and ATLAS~\cite{Aad:2015vsa}. However, there are some significant kinematic differences. The primary difference is that the muon (electron) in the LFV $\PH \to \PGm(\Pe) \PGt$ decay is produced promptly, and tends to have a higher momentum than in the $\PH \to \PGt_{\PGm(\Pe)}\PGt$ decay. Only the gluon fusion production process is considered in this search and the signal is modelled assuming a narrow width of the Higgs boson. The strategy is similar to the previous LFV $\PH(125)$ searches performed by the CMS Collaboration, but optimised for higher mass Higgs boson decays.

This paper is organized as follows. After a brief overview of the CMS detector in Section~\ref{cmsdet} and the description of the collision data and simulated samples used in the analysis in Section~\ref{samples}, the event reconstruction is described in Section~\ref{reconst}. The event selection is outlined in Section~\ref{eventsel} and the background processes are described in Section~\ref{backgrounds}. This is followed by a description of the systematic uncertainties in Section~\ref{systematics}. Finally, the results are presented in Section~\ref{results},  and the paper is summarized in Section~\ref{sec:summary}.

\section{The CMS detector\label{cmsdet}}

A detailed description of the CMS detector, together with a definition of the coordinate system used and the relevant kinematic variables, can be found in Ref.~\cite{CMS-JINST}. The central feature of the CMS apparatus is a superconducting solenoid of 6\unit{m} internal diameter, providing a magnetic field of 3.8\unit{T}. Within the solenoid volume are a silicon pixel and strip tracker, a lead tungstate crystal electromagnetic calorimeter (ECAL), and a brass and scintillator hadron calorimeter (HCAL), each composed of a barrel and two endcap sections. Forward calorimeters extend the pseudorapidity ($\eta$) coverage provided by the barrel and endcap detectors. Muons are detected in gas-ionization chambers embedded in the steel flux-return yoke outside the solenoid. Events of interest are selected using a two-tiered trigger system~\cite{Khachatryan:2016bia}. The first level, composed of custom hardware processors, uses information from the calorimeters and muon detectors to select events at a rate of around 100\unit{kHz} within a time interval of less than 4\mus. The second level, known as the high-level trigger, consists of a farm of processors running a version of the full event reconstruction software optimised for fast processing, and reduces the event rate to around 1\unit{kHz} before data storage.

\section{Collision data and event simulation \label{samples}}

The data used in this analysis have been collected in $\Pp\Pp$ collisions at the LHC, at a centre-of-mass energy of $13$\TeV, with the CMS detector in 2016, and correspond to an integrated luminosity of 35.9\fbinv~\cite{CMS-PAS-LUM-17-001}. A trigger requiring at least one muon is used to collect the data sample in the $\PH \to \PGm\tauh$ and $\PH \to \PGm\PGt_{\Pe}$ channels. Triggers requiring at least one electron, or a combination of an electron and a muon are used for the $\PH \to \Pe \tauh$ and $\PH \to \Pe \PGt_{\PGm}$ channels respectively.
Simulated samples of signal and background events are produced with different event generators. The $\PH \to \PGm \PGt$ and $\PH \to \Pe \PGt$ decay samples are generated with \POWHEG 2.0~\cite{Nason:2004rx,Frixione:2007vw, Alioli:2010xd, Alioli:2010xa, Alioli:2008tz, Bagnaschi:2011tu} at next-to-leading-order (NLO) in perturbative quantum chromodynamics. Only the gluon fusion ($\Pg\Pg\PH$)~\cite{Georgi:1977gs} production mode has been considered in this analysis. These scalar boson samples are generated assuming the narrow width approximation for a range of masses from 200 to 900\GeV.
The $\zjets$ and $\wjets$ processes are simulated using the \MGvATNLO 2.2.2~\cite{Alwall:2014} generator at leading order with the MLM jet matching and merging~\cite{Alwall:2007fs}. The \MGvATNLO generator is also used for diboson production which is simulated at  NLO with the FxFx jet matching and merging scheme~\cite{Frederix:2012ps}. The $\POWHEG$ 2.0 and 1.0 at NLO are used for top quark-antiquark ($\ttbar$) and single top quark production, respectively.
The \POWHEG and  \MGvATNLO generators are interfaced with \PYTHIA~8.212 ~\cite{Sjostrand:2014zea} for parton showering and fragmentation.
The \PYTHIA parameters for the underlying event description are set
to the {CUETP8M1} tune~\cite{Khachatryan:2015pea}. The set of parton distribution functions (PDFs) used is NNPDF30nloas0118~\cite{BALL2012153}. The CMS detector response is modelled using \GEANTfour~\cite{GEANT4}.

Because of the high instantaneous luminosities attained during data taking, events have multiple $\Pp\Pp$ interactions per bunch crossing (pileup). This effect is taken into account in simulated samples, by generating concurrent minimum bias events, and overlapping them with simulated hard events. All simulated samples are weighted to match the pileup distribution observed in data, which has an average of approximately 23 interactions per bunch crossing.

\section{Event reconstruction \label{reconst}}

The event reconstruction is performed using a particle-flow (PF) algorithm, which aims to reconstruct and identify each individual particle in an event (PF candidate), with an optimised combination of information from the various elements of the CMS detector~\cite{Sirunyan:2017ulk}.
In this process, the identification of the particle type for each PF candidate (photon, electron, muon, charged or  neutral hadron) plays an important role in the determination of the particle direction and energy. The primary $\Pp\Pp$  vertex of the event is identified as the reconstructed vertex with the largest value of summed physics-object $\pt^2$, where \pt is the transverse momentum. The physics objects are the jets, clustered using the jet finding algorithm~\cite{Cacciari:2008gp,Cacciari:2011ma} with the tracks assigned to the vertex as inputs, and the associated missing transverse momentum, taken as the negative vector sum of the \pt of those jets.

A muon is identified as a track in the silicon detectors, consistent with the primary
$\Pp\Pp$  vertex and with either a track or several hits in the muon system, associated
with an energy deposit in the calorimeters compatible with the expectations for a
muon~\cite{Chatrchyan:2018xi,Sirunyan:2017ulk}. Identification is based on the number of
spatial points measured in the tracker and in the muon system, the track quality, and its
consistency with the event vertex location. The identification working point chosen for
this analysis reconstructs muons with an efficiency above 98\% and a hadron
misidentification rate of 0.1\% for pions and 0.3\% for kaons. 
The energy is obtained from the corresponding track momentum. An important aspect of muon reconstruction is the lepton isolation that is described later in this section.

An electron is identified as a  charged-particle track from the primary $\Pp\Pp$  vertex in combination with one or more ECAL energy clusters. These clusters  are matched with the track extrapolation to the ECAL and with  possible bremsstrahlung photons emitted when interacting with the material of the tracker~\cite{Khachatryan:2015hwa}. Electron candidates are accepted in the pseudorapidity range $\abs{\eta}<2.5$, with the exception of the region $1.44<\abs{\eta}<1.57$ where service infrastructure for the detector is located. They are identified using a multivariate-analysis (MVA) discriminator  that combines observables sensitive to the amount of bremsstrahlung along the electron trajectory, the geometric and momentum matching between the electron trajectory and associated clusters, as well as various shower shape observables in the calorimeters. Electrons from photon conversions are removed. The chosen working point for selecting the electrons assures an average identification efficiency of 80\% with a misidentification probability of ~5\%.
The energy of electrons is determined from a combination of the track momentum at the primary vertex, the corresponding ECAL cluster energy, and the energy sum of all bremsstrahlung photons associated with the track.

Charged hadrons are identified as charged-particle tracks neither identified as electrons, nor as muons. Finally, neutral hadrons are identified as HCAL energy clusters not linked to any charged hadron trajectory, or as a combined ECAL and HCAL energy excess with respect to the expected charged hadron energy deposit.  All the PF candidates are clustered into hadronic jets
using the infrared- and collinear-safe anti-\kt algorithm~\cite{Cacciari:2008gp}, implemented in the \FASTJET package~\cite{Cacciari:fastjet},  with a distance parameter of 0.4.
Jet momentum is determined as the vectorial sum of all particle momenta in the jet, and is found from simulation to be, on average, within 5 to 10\% of the true momentum over the whole \pt spectrum and detector acceptance. Additional proton-proton interactions within the same or nearby bunch crossings can contribute additional tracks and calorimetric energy depositions, increasing the apparent jet momentum. To mitigate this effect, tracks identified to be originating from pileup vertices are discarded and an offset correction is applied to correct for remaining contributions. Jet energy corrections are derived from simulation studies so that the average measured response of jets becomes identical to that of particle level jets. In situ measurements of the momentum balance in dijet, photon+jet, \zjets, and multijet events are used to determine any residual differences between the jet energy scale in data and in simulation, and appropriate corrections are made~\cite{Khachatryan:2016kdb}. Additional selection criteria are applied to each jet to remove jets potentially dominated by instrumental effects or reconstruction failures.

Hadronically decaying $\PGt$ leptons (\tauh) are reconstructed and identified using the hadrons-plus-strips algorithm~\cite{Khachatryan:2015dfa,Sirunyan:tau16}.
The reconstruction starts from a jet and searches for the products of the main $\PGt$ lepton decay modes: one charged hadron and up to two neutral pions, or three charged hadrons. To improve the reconstruction efficiency in the case of conversion of the photons from a neutral-pion decay, the algorithm considers the PF photons and electrons from a strip along $\phi$. The sign of the \tauh candidate is determined through its decay products.

An MVA discriminator, based on variables such as lifetime information, decay mode, multiplicity of neutral, charged and pileup particles  in a cone around the reconstructed \tauh, is used to reduce the rate for quark- and gluon-initiated  jets identified as \tauh candidates. The working point used in the analysis is a ``tight" one, with an efficiency of about 50\% for a genuine \tauh, and approximately a 0.2\% misidentification rate for quark and gluon jets~\cite{Sirunyan:tau16}. Additionally, muons and electrons misidentified as \tauh are rejected by  considering the consistency between the measurements in the tracker, calorimeters, and muon detectors.
The specific identification criteria depend on the final state studied and
on the background composition. The $\PGt$ leptons that decay to muons and electrons are reconstructed
in the same manner as prompt muons and electrons, respectively, as described above.

The variable $\Delta R = \sqrt{\smash[b]{(\Delta\eta)^2 +(\Delta\phi)^2}}$
is used to measure the separation between reconstructed objects in the detector, where $\eta$ and $\phi$ are the pseudorapidity and azimuthal directions, respectively.

Jets misidentified as muons or electrons are suppressed by imposing  isolation requirements.
The muon (electron) isolation is measured relative to its $\pt^{\ell}$ ($\ell = \PGm, \Pe$) by summing over the $\pt$ of PF particles
in a cone with $\Delta R = 0.4$ (0.3) around the lepton, excluding the lepton itself:
\begin{linenomath*}
  \begin{equation*}
    I_\text{rel}^{\ell} = \frac{ \sum  \pt^\text{charged} + \text{max}\left[ 0, \sum \pt^\text{neutral}
                                 +  \sum \pt^{\gamma} - \pt^\text{PU}\left(\ell\right)  \right] }{ \pt^{\ell}},
  \end{equation*}
\end{linenomath*}
where $\pt^\text{charged}$, $\pt^\text{neutral}$, and $\pt^{\gamma}$  indicate the \pt of a charged and of a neutral particle, and a photon within the cone, respectively.
The neutral particle contribution to isolation from pileup, $\pt^\text{PU}\left(\ell\right)$, is estimated from the $\pt$
sum of charged hadrons not originating
from the primary vertex scaled by a factor of 0.5~\cite{Chatrchyan:2018xi} for the muons.
For the electrons, this contribution is estimated  from the area of the jet and the average energy density of the event~\cite{1126-6708-2008-04-005, CACCIARI2008119}.
The charged-particle contribution to isolation from pileup is rejected by requiring the tracks to originate from the primary vertex.
Jet arising from a \PQb quark are identified by the combined secondary vertex \PQb tagging algorithm~\cite{Sirunyan:2017ezt} using the working point characterised by a \PQb jet identification efficiency around 65\% and  a misidentification probability around 1\% for light quark and gluon jets.

All the reconstructed particles in the event are used to estimate the missing transverse momentum, $\ptvecmiss$, which is defined as the projection onto the plane perpendicular to the beam axis of the negative vector sum of the momenta of all reconstructed PF candidates in an event~\cite{Sirunyan:2019kia}. The effect of the jet energy corrections described earlier in this section is then propagated to this $\ptvecmiss$.
The magnitude of the final vector is referred to as \ptmiss. The transverse mass $\mT(\ell)$ is a variable formed from the lepton transverse momentum and the missing transverse momentum vectors:
$ {\mT}(\ell)=\sqrt{\smash[b]{2|\ptvec^{\ell}||\ptvecmiss|(1-\cos{\Delta \phi_{\ell - \ptmiss}})}} $, where $\Delta \phi_{\ell - \ptmiss}$ is the angle between the lepton transverse momentum and the missing transverse momentum.
The collinear mass, $\mcol$, provides an estimate of $m_{\PH}$ using the observed
decay products of the Higgs boson candidate. It is reconstructed using the collinear approximation based on
the observation that, since $m_{\PH}\gg m_{\PGt}$, the $\PGt$ lepton decay products are
highly boosted in the direction of the  $\PGt$ candidate~\cite{Ellis:1987xu}.
The neutrino momenta can be approximated to have the same
direction as the other visible decay products of the $\PGt$ lepton ($\vec{\PGt}^\text{vis}$)
and the component of the $\ptvecmiss$ in the direction of the visible $\PGt$ lepton  decay products is used to
estimate the transverse component of the neutrino momentum ($\pt^{\nu,\,\text{est}}$). The collinear
mass is then $\mcol= M_{\text{vis}} / \sqrt{\smash[b]{x_{\PGt}^\text{vis}}}$, where $x_{\PGt}^\text{vis}$ is the
fraction of momentum carried by the visible decay products of the $\PGt$ lepton,
$x_{\PGt}^\text{vis}={\pt^{\vec{\PGt}^{\text{vis}}}}/{(\pt^{\vec{\PGt}^\text{vis}}+\pt^{\nu,\,\text{est}})}$,
and $M_{\text{vis}}$ is the visible mass of the $\PGt-\Pe$ or $\PGt-\PGm$ system.

Dedicated performance studies on data validate the reconstruction and identification
techniques described in this section. When necessary, corrections
have been applied to the simulated samples to ensure they correctly describe the behaviour
of the
data~\cite{Chatrchyan:2018xi,Khachatryan:2015hwa,Khachatryan:2016kdb,Sirunyan:tau16,Sirunyan:2017ezt,Sirunyan:2019kia}.

\section{Event selection \label{eventsel}}
The event selection is performed in two steps. An initial selection is followed by another, final, set of requirements on kinematic variables that exploit the distinct event topology of the signal. The event sample defined by the initial selection is used in the background estimation described in Section~\ref{backgrounds}. The event selection begins by requiring two isolated leptons of opposite charge, different flavour, and separated by $\Delta R > 0.3$. The isolation of the \tauh candidates is included in the MVA discriminator described in Section~\ref{reconst}. Events with additional \PGm, \Pe, or $\tauh$ candidates respectively with $\pt > 10$, 5, or 20\GeV are discarded. The kinematic requirements applied are dictated by the triggers or detector acceptance and are summarized in Table~\ref{tab:loose_sel}.

The events are then divided into two categories according to the number of jets in the event. The jets are required to have $\pt>30$\GeV and $\abs{\eta}<4.7$. Events with no jets form the 0-jet category while events with exactly one jet form the 1-jet category. The 1-jet category includes  $\Pg\Pg\PH$ production with initial state radiation. Events with more than one jet are discarded.

\begin{table}[htpb]
 \centering
 \topcaption{Initial selection criteria applied to the kinematic variables for the \Hmt\ and \Het\ analyses. The selected sample is used in the background estimation from control samples in data.}
  \begin{tabular}{ccccc} \hline
    & $~\Hmuhad~$ &  $~\Hmue~$  & $~\Hehad~$  &  $~\Hemu~$   \\ \hline
    $\pt^{\PGm}$ & $>$53\GeV &  $>$53\GeV&\NA &$>$10\GeV      \\
    $\pt^{\Pe}$    &\NA& $>$10\GeV & $>$26\GeV & $>$26\GeV \\
    $\pt^{\PGt}$   & $>$30\GeV &\NA &$>$30\GeV&\NA \\
    $\abs{\eta^{\PGm}}$ & $<$2.4 &  $<$2.4 & \NA &$<$2.4                                       \\
    $\abs{\eta^{\Pe}} $ &\NA&$<$2.4 & $<$2.1 & $<$2.4   \\
    $\abs{\eta^{\PGt}} $ &$<$2.3  &\NA  &$<$2.3 &\NA        \\
    $I_{\text{rel}}^{\PGm}$  & 0.15 & $<$0.15 &  \NA & $<$0.15                                \\
    $I_{\text{rel}}^{\Pe}$  & \NA & $<$0.1 &  $<$0.1 & $<$0.1  \\
    $\Delta R(\PGm,\Pe)$  &\NA&$>$0.3&\NA&$>$0.3\\
    $\Delta R(\PGm,\PGt)$  &$>$0.3&\NA&\NA&\NA\\
    $\Delta R(\Pe,\PGt)$   &\NA&\NA&$>$0.3&\NA\\
    \hline
  \end{tabular}
  \label{tab:loose_sel}
\end{table}

The final selection is given in Table~\ref{tab:final_sel}. It begins by tightening the \pt requirement of the prompt lepton from the Higgs boson decay, as it provides a powerful discriminant against the background. The $\PGt$ lepton in the \PH decay is highly  boosted, leading to a collimation of the decay products.
This can be exploited by either limiting the  azimuthal separation of the decay products including the \ptvecmiss, or imposing a  requirement on the  transverse mass $\mT(\PGt)$, which is strongly correlated with the azimuthal separation. These selection criteria are optimised for each decay mode in two $m_{\PH}$ ranges to obtain the most stringent expected upper limits. The low- and high-mass regions are defined to be $200<m_{\PH}<450$\GeV and $450<m_{\PH}<900$\GeV, respectively.
A binned likelihood fit to the \mcol distributions is then used to extract signal and
background contributions.  The \mcol approximates the Higgs mass better than the widely
used \mvis, and therefore improves the separation of the signal from the background. This
improvement is larger in the high mass regime, with up to a factor of three
gain in sensitivity when compared to the use of $M_{\text{vis}}$.

\begin{table}[hbtp]
  \centering
  \topcaption{Final event selection criteria for the low-mass range, $200<m_{\PH}<450$\GeV, and the high-mass range, $450<m_{\PH}<900$\GeV, considered in the \Hmt\ and \Het\ analyses.}
  \begin{tabular}{cccccc}
  \hline
  &&& Low-mass range && High-mass range\\ \hline
\multirow{6}{*}{\Hmuhad}   & \multirow{3}{*}{0-jet} && $\pt^{\PGm}>60$\GeV &&  $\pt^{\PGm}>150$\GeV\\
  & && $\pt^{\PGt}>30$\GeV &&  $\pt^{\PGt}>45$\GeV\\
  & && $\mT(\tauh)<105$\GeV &&  $\mT(\tauh)<200$\GeV \medskip\\
 &  \multirow{3}{*}{1-jet}  && $\pt^{\PGm}>60$\GeV &&  $\pt^{\PGm}>150$\GeV\\
 & && $\pt^{\PGt}>30$\GeV &&  $\pt^{\PGt}>45$\GeV\\
 &  && $\mT(\tauh)<120$\GeV &&  $\mT(\tauh)<230$\GeV \\
    [\cmsTabSkip]
 \multirow{8}{*}{\Hmue} &\multirow{4}{*}{0-jet}  && $\pt^{\PGm}>60$\GeV &&  $\pt^{\PGm}>150$\GeV\\
 & &&  $\pt^{\Pe}>10$\GeV &&  $\pt^{\Pe}>10$\GeV\\
 &  && $\Delta\phi(\Pe, \ptvecmiss)<0.7$\unit{rad} && $\Delta\phi(\Pe, \ptvecmiss)<0.3$\unit{rad} \\
&  && $\Delta\phi(\Pe, \PGm)>2.2$\unit{rad} && $\Delta\phi(\Pe, \PGm)>2.2$\unit{rad} \medskip \\
  &  \multirow{4}{*}{1-jet}  && $\pt^{\PGm}>60$\GeV && $\pt^{\PGm}>150$\GeV\\
&  && $\pt^{\Pe}>10$\GeV && $\pt^{\Pe}>10$\GeV\\
&  && $\Delta\phi(\Pe, \ptvecmiss)<0.7$\unit{rad} && $\Delta\phi(\Pe, \ptvecmiss)<0.3$\unit{rad}\\
  &  && $\Delta\phi(\Pe, \PGm)>2.2$\unit{rad}&& $\Delta\phi(\Pe, \PGm)>2.2$\unit{rad}\\
    [\cmsTabSkip]
 \multirow{6}{*}{\Hehad} &\multirow{2}{*}{0-jet}  && $\pt^{\Pe}>60$\GeV &&  $\pt^{\Pe}>150$\GeV\\
&  && $\pt^{\PGt}>30$\GeV &&  $\pt^{\PGt}>45$\GeV\\
&  && $\mT(\tauh)<105$\GeV &&  $\mT(\tauh)<200$\GeV\medskip\\
&  \multirow{2}{*}{1-jet}  && $\pt^{\Pe}>60$\GeV &&  $\pt^{\Pe}>150$\GeV\\
&  && $\pt^{\PGt}>30$\GeV &&  $\pt^{\PGt}>45$\GeV\\
  &  && $\mT(\tauh)<120$\GeV &&  $\mT(\tauh)<230$\GeV\\
    [\cmsTabSkip]
\multirow{8}{*}{\Hemu} & \multirow{3}{*}{0-jet}  && $\pt^{\Pe}>60$\GeV &&  $\pt^{\Pe}>150$\GeV\\
& && $\pt^{\PGm}>10$\GeV &&  $\pt^{\PGm}>10$\GeV\\
&  && $\Delta\phi(\PGm, \ptvecmiss)<0.7$\unit{rad} && $\Delta\phi(\PGm, \ptvecmiss)<0.3$\unit{rad} \\
&  && $\Delta\phi(\Pe, \PGm)>2.2$\unit{rad} && $\Delta\phi(\Pe, \PGm)>2.2$\unit{rad} \medskip\\
&  \multirow{3}{*}{1-jet}  && $\pt^{\Pe}>60$\GeV && $\pt^{\Pe}>150$\GeV\\
& && $\pt^{\PGm}>10$\GeV &&$\pt^{\PGm}>10$\GeV\\
&  && $\Delta\phi(\PGm, \ptvecmiss)<0.7$\unit{rad} && $\Delta\phi(\PGm, \ptvecmiss)<0.3$\unit{rad}\\
&  && $\Delta\phi(\Pe, \PGm)>2.2$\unit{rad}&& $\Delta\phi(\Pe, \PGm)>2.2$\unit{rad}\\ \hline
  \end{tabular}
  \label{tab:final_sel}
\end{table}

\section{Background estimation}
\label{backgrounds}

The most significant  background in the \muhad and \ehad channels comes from the $\wjets$ process and from events comprised uniquely of jets  produced through the strong interaction, referred to as quantum chromodynamics (QCD) multijet events. In these processes, jets are misidentified as electrons, muons or \PGt leptons. This  background is estimated with the collected data. The main background in the  \mue and \emu channels is $\ttbar$ production. It is estimated using simulations. 	
Other smaller backgrounds include electroweak diboson ($\PW\PW$, $\PW\PZ$, and $\PZ\PZ$),
Drell-Yan (DY)$\to\ell\ell$ ($\ell = \Pe, \PGm)+\text{jets}$, DY$\to\PGt\PGt+\text{jets}$,
SM Higgs boson ($\PH \to \PGt\PGt,\PW\PW$), $\PW\gamma^{(*)}\!+\!\text{jets}$, and single
top quark production processes. These are estimated using simulations. Gluon fusion,
vector boson fusion, and associated production mechanisms are considered for the SM Higgs boson background. The background estimation techniques are described in detail below, and are validated with control regions that are enhanced with the dominant backgrounds.

The DY$\to\ell\ell,\PGt\PGt$ background is estimated from simulation. A reweighting is applied to the generator-level \PZ boson $\pt$ and invariant mass,  $m_{\ell\ell,\PGt\PGt}$, distributions to correct for a shape discrepancy between data and simulation. The reweighting factors, extracted from a control region enriched in $\PZ\to\PGm\PGm$ events, are applied in bins of \PZ boson $\pt$ and $m_{\ell\ell,\PGt\PGt}$ as explained in \cite{CMS-PAS-HIG-16-043}. Additional corrections for $\Pe\to\tauh$ and $\mu\to\tauh$ misidentification rates are applied to the simulated DY sample when the reconstructed $\tauh$ candidate is matched to an electron for the $\Hehad$ channel or a muon for the $\Hmuhad$ channel, respectively, at the generator level. These corrections depend on the lepton $\eta$ and are measured in $Z\to\ell\ell$ data events.

\begin{figure}[htpb]
\centering
 {
\includegraphics[width=0.49\textwidth]{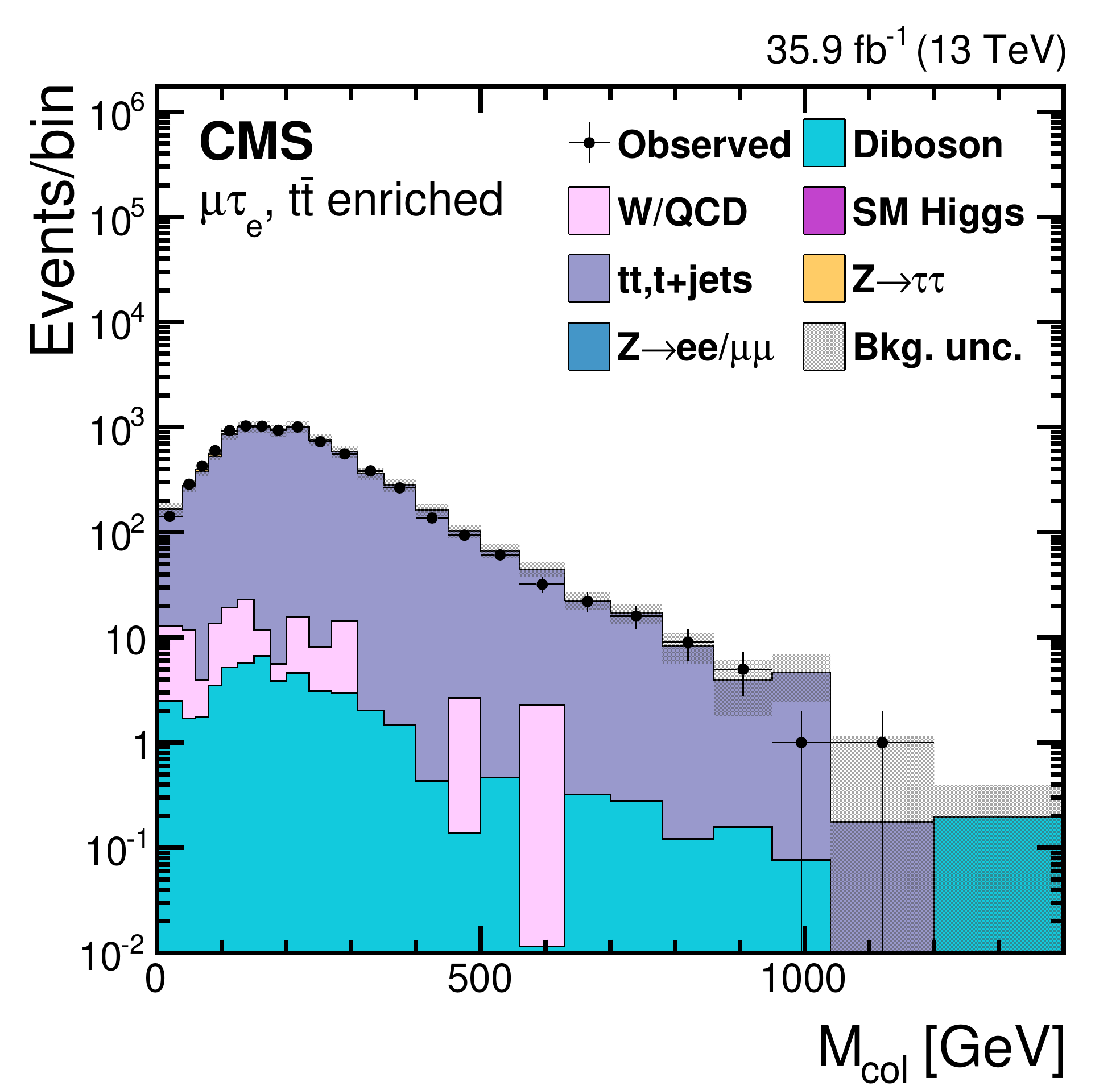}
\includegraphics[width=0.49\textwidth]{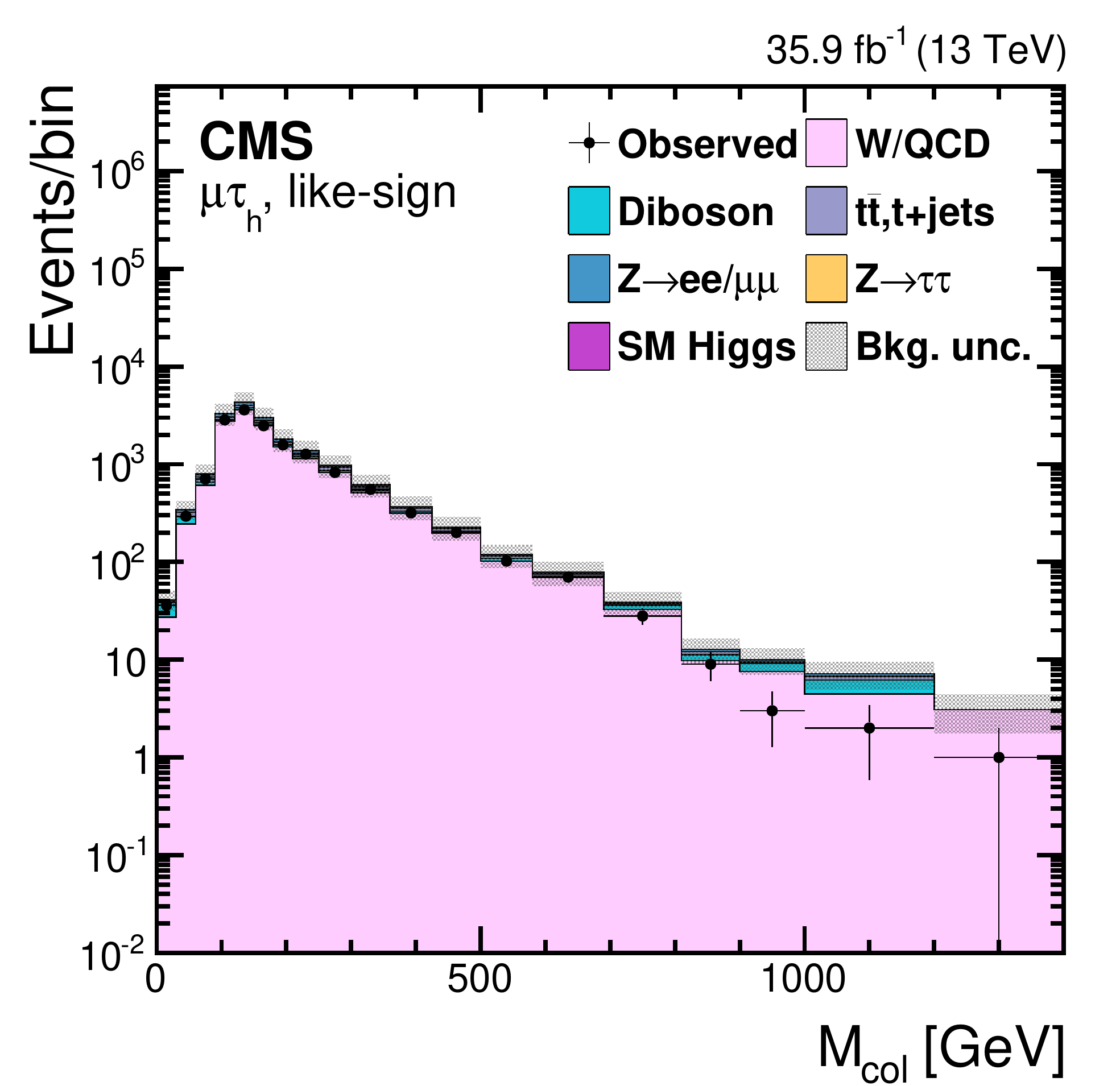}
\includegraphics[width=0.49\textwidth]{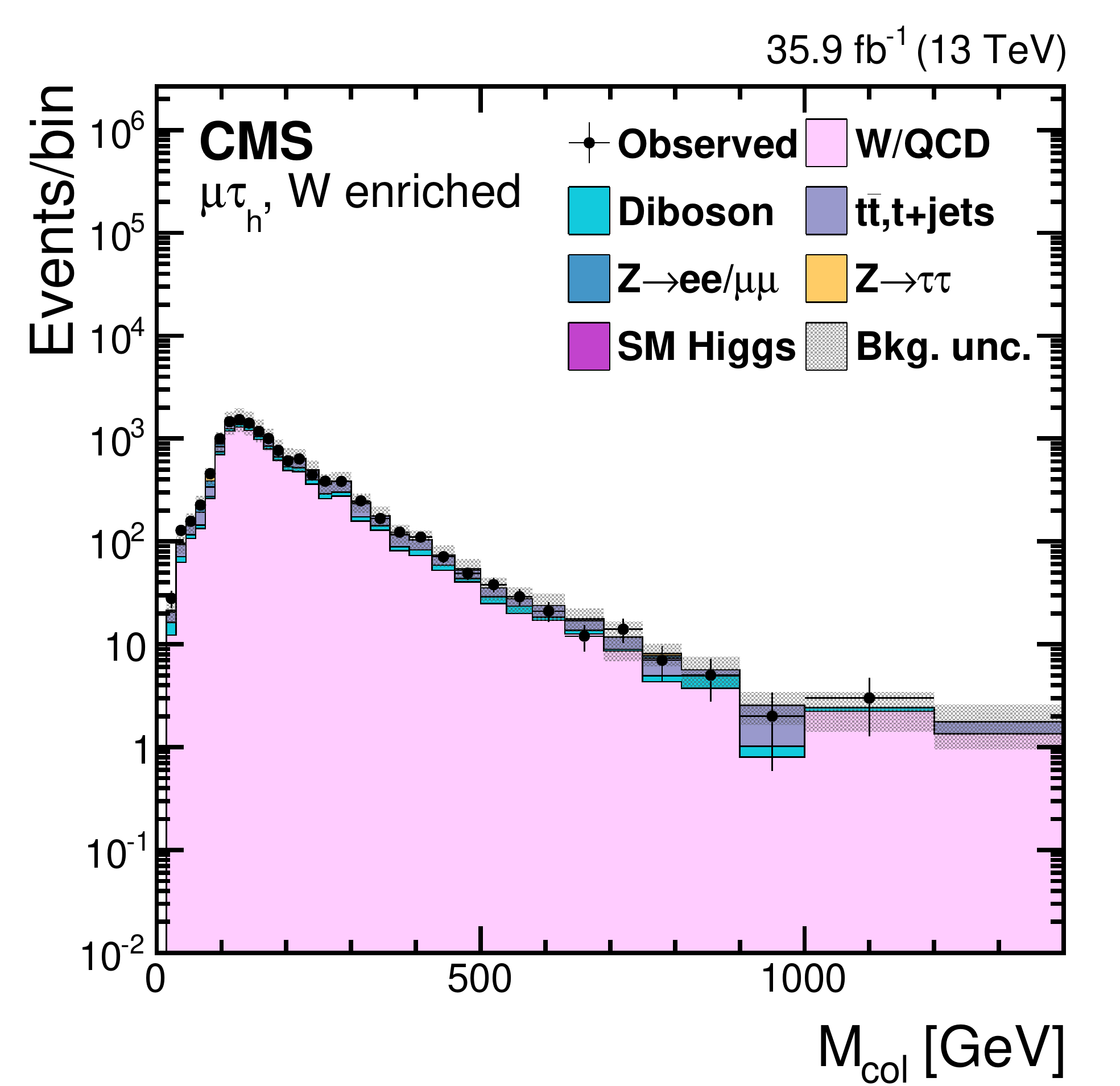}
}\\
\caption{The \mcol distribution in the $\ttbar$ enriched (upper left), like-sign lepton
(upper right), and $\wjets$ enriched (lower) control samples defined in the text. The
uncertainty bands include both statistical and systematic uncertainties from
Section~\ref{systematics}. No fit is performed for these distributions. The different
background processes shown are normalised to the luminosity of the data either using the
theoretical prediction of the corresponding production cross section or directly from the data driven
technique described in the text.}
\label{fig:SScontrolregion}
\end{figure}

The $\ttbar$ background is also estimated using simulation. The overall normalisation of this estimate in the signal region is corrected with a rescaling factor derived from a control region enriched in $\ttbar$~events, defined by requiring the initial selection with the additional requirement that at least one of the jets is \PQb tagged. Figure~\ref{fig:SScontrolregion} (upper left) shows the data compared to the background estimate in the $\ttbar$-enriched region in the \Hmue channel.

Jets from $\wjets$ and QCD multijet events that are misidentified as electrons, muons and, mainly, \PGt leptons, are leading source of background in the \muhad and \ehad channels. In $\wjets$ events, one lepton candidate is expected to be a genuine lepton from the $\PW$ decay and the
other a jet misidentified as a lepton. In QCD multijet events, both lepton candidates
are misidentified jets. A technique fully based on control samples in data is used to estimate the misidentified lepton background in the \muhad and \ehad channels, for which it is the  dominant contribution. In the \mue and \emu channels, this background is estimated using a combination of simulated samples and control regions in data. These methods have been used in Refs.~\cite{HIG-17-001} and~\cite{CMS-PAS-HIG-16-043}, and  a detailed description can be found in those publications. However, we are briefly describing the techniques in the following subsections.

\subsection{Misidentified lepton background estimation from control samples in data}

The misidentified-lepton background is estimated from data. The misidentification
probabilities, $f_{i}$, where $i=\PGm$, $\Pe$, or $\tauh$,  are evaluated with independent
\zjets data sets and then applied to a control sample. The control sample is obtained by
relaxing the signal selection requirements, the  $\PGm$, $\Pe$, or $\tauh$ isolation, and excluding events passing the signal selection.
The $f_{i}$ are estimated using events with a \PZ boson candidate and one jet that can be misidentified as \PGm, \Pe, or \PGt. The \PZ boson candidate is formed requiring two muons with  $\pt>26$\GeV, $\abs{\eta}<2.4$, and $I_\text{rel}^{\PGm}<0.15$. The muons are required to have opposite charges and the dimuon invariant mass, $m_{\PGm\PGm}$,  must satisfy  $76<m_{\PGm\PGm}<106$\GeV. The contribution from  diboson events, where the third lepton candidate corresponds to a genuine \PGm, \Pe, or \PGt, is subtracted using
simulation. Two \zjets samples are defined: a signal-like one, in which the jet satisfies the same \PGm, \Pe, or \PGt selection criteria used in the $\PH \to \PGm\PGt$ or $\PH \to \Pe\PGt$ selections,  and a background-enriched \zjets sample with relaxed  identification on the jet misidentified as \PGm, \Pe, or \PGt, but excluding events selected in the signal-like sample. The requirements on the third candidate, \ie the misidentified jet,  depend  on the lepton flavour. The two samples are used to estimate $f_{i}$ as
\begin{linenomath*}
  \begin{equation*}
    f_{i}=\frac{N_{i}(\text{\zjets signal-like})}{N_{i}(\text{\zjets background-enriched})+N_{i}(\text{\zjets signal-like})},
  \end{equation*}
\end{linenomath*}
where  $N_{i}(\text{\zjets signal-like})$ is the number of events with a third  candidate (\PGm, \Pe, or \PGt) that passes the signal-like sample selection, and $N_{i}(\text{\zjets background-enriched})$ is the number of events in the background-enriched sample. The background-enriched selection used to estimate  the misidentified $\PGm$ and $\Pe$ contribution requires an isolation of $0.15<I_\text{rel}^{\PGm}<0.25 $ and $0.1<I_\text{rel}^{\Pe}<0.5$, respectively. In both cases the misidentification rate is computed and applied as a function of the lepton \pt. The lepton selection for the \tauh\ background-enriched sample requires that the \tauh lepton candidates are identified using a loose $\tauh$ identification and isolation working point but are not identified by the tight working point used for the signal selection. The loose and tight working points have an efficiencies of 70 and 50\% for genuine \tauh\ candidates, respectively.

The \tauh misidentification rates have a \pt dependence that varies with the number of charged pions in the  decay. They are estimated and applied  as a function of \pt and for either one or three charged pions in the decay.
The misidentified background in the signal sample is obtained from  control samples for each lepton flavour.  The selection requirements for these samples are the same as for the signal sample except that the \PGm, \Pe, or \PGt should pass the identification and isolation criteria used for the \zjets background-enriched sample, but not those defining the \zjets signal-like sample. To estimate the misidentified background in the signal sample, each  event in this background enriched sample  is weighted by a factor ${f_{i}}/{(1-f_{i})}$.
The background from misidentified muons and electrons is estimated  to be less than 5\% of the misidentified $\tauh$ lepton background and is neglected.

The background estimate is validated in a like-sign sample by applying  the misidentification rate $f_{i}$ to events selected by requiring the \PGm, \Pe, or \PGt in the pairs having the same charge in both the back\-ground-\-en\-riched and the signal-like samples. This validation is performed after the initial selection described in Section~\ref{eventsel}.  Figure~\ref{fig:SScontrolregion} (upper right) shows the data compared to the background estimate in the like-sign control region for the \Hmuhad channel. The like-sign selection enhances the misidentified-lepton background, and this sample is expected to be composed of a similar fraction of \wjets and QCD multijet events. The background estimate is also validated in a \PW\ boson enriched control sample. This data sample is obtained by applying the signal sample requirements and \mT cuts,  $50<\mT(\ell)<110$\GeV ($\ell = \PGm$  or \Pe) and $\mT(\tau)>50$\GeV. The misidentified background in the signal region and \PW\ boson enriched control sample are both dominated by \wjets events, with QCD multijet events forming a small fraction of the samples. Figure~\ref{fig:SScontrolregion} (lower) shows the data compared to the background estimate in the \wjets enriched sample for the \Hmuhad channel. The background expectation for the \Hehad channel is also validated with the same samples and gives similar agreement.

 \subsection{\texorpdfstring{\wjets}{W+jets} and QCD background estimation in \texorpdfstring{$\mu\tau_{\Pe}$}{mu taue} and \texorpdfstring{$\Pe\tau_{\mu}$}{e taumu} channels}
The \wjets background contribution to the misidentified background is estimated with
simulations. The QCD multijet contribution is estimated with like-sign data events that pass all the other signal requirements. The remaining non-QCD background is estimated using simulation. The resulting sample is then rescaled to account for the differences
between the background composition in the like and opposite sign samples. The scaling
factors are extracted from QCD multijet enriched control samples, composed of events where
the lepton candidates satisfy inverted isolation requirements, as explained in
Ref.~\cite{CMS-PAS-HIG-16-043}. This background contribution accounts for a negligible
fraction of the total yield after selection in both $\mu\tau_{\Pe}$ and $\Pe\tau_{\mu}$ channels. 

\section{Systematic uncertainties}\label{systematics}
Systematic uncertainties arise from both experimental and theoretical sources and can affect the normalisation and the shape of the collinear mass distribution. They are summarized in Table~\ref{tab:systematics}.
\begin{table}[htpb]
\centering
  \topcaption{The systematic uncertainties for the four channels. All uncertainties are treated as correlated between the categories, except those with more values separated by the $\oplus$~symbol. In the case of two values, the first value is the correlated uncertainty and the second value is the uncorrelated uncertainty for each individual category. In the case of three values, the first and second values correspond to the uncertainties arising from factorisation and renormalisation scales and PDF variations and are correlated between categories, while the third value is the uncorrelated uncertainty for each individual category. Two values separated by the ``--" sign represent the range of the uncertainties from the different sources and/or in the different jet categories.
}
\label{tab:systematics}
\cmsTable{
\begin{tabular}{l*{4}{c}}
Systematic  uncertainty            & $\PH\to\PGm\tauh$& $\PH\to\PGm\PGt_{\Pe}$ & $\PH\to\Pe \tauh$    & $\PH\to\Pe\PGt_{\PGm}$ \\ \hline
Muon  trigger/ID/isolation         &       2\%             &       2\%          &         \NA         &  2\%             \\
Electron trigger/ID/isolation      &      \NA            &       2\%          &          2\%          &    2\%            \\
Hadronic $\tauh$ efficiency         &       5\%             &      \NA         &          5\%          & \NA                \\
High $\pt$ $\tauh$  efficiency      & $^{+5}_{-35}\%{\times}\pt{\times}0.001$ & \NA & $^{+5}_{-35}\%{\times}\pt{\times}0.001$ & \NA                \\
\PQb tagging veto                     &      2.0--2.5\%       &      2.0--2.5\%    &       2.0--2.5\%      &  2.0--2.5\%          \\
[\cmsTabSkip]
$\PGm\to\tauh$ background         &           25\%        &       \NA        &         \NA         & \NA                             \\
$\Pe\to\tauh$ background          &          \NA        &         \NA      &           12\%        & \NA                              \\
$\text{jet}\to\tauh$ background &     30\%$\oplus$10\%  &          \NA     &    30\%$\oplus$10\%   & \NA                               \\
QCD multijet background            &     \NA             &       30\%         &         \NA         &      30\%                 \\
$\PZ\to\mu\mu/\Pe\Pe+\textrm{jets}$ background&    \NA              & 0.1\%$\oplus$2\%$\oplus$5\%    &       \NA           &     0.1\%$\oplus$2\%$\oplus$5\%    \\
$\PZ\to\tau\tau+\text{jets}$ background &     0.1\%$\oplus$2\%$\oplus$5\%   &   0.1\%$\oplus$2\%$\oplus$5\%  &     0.1\%$\oplus$2\%$\oplus$5\%   & 0.1\%$\oplus$2\%$\oplus$5\%         \\
$\wjets$ background     &    \NA              &  0.8\%$\oplus$3.8\%$\oplus$5\%              &         \NA         &      0.8\%$\oplus$3.8\%$\oplus$5\%                \\
$\PW\PW,\PZ\PZ,\PW\PZ$ background&     3.5\%$\oplus$5\%$\oplus$5\%    &   3.5\%$\oplus$5\%$\oplus$5\%   &     3.5\%$\oplus$5\%$\oplus$5\%    & 3.5\%$\oplus$5\%$\oplus$5\%             \\
$\PW\!+\!\gamma$ background           &    \NA              &   10\%$\oplus$5\%  &         \NA         &   10\%$\oplus$5\%            \\
Single top quark background        &     3\%$\oplus$5\%$\oplus$5\%    &   3\%$\oplus$5\%$\oplus$5\%   &     3\%$\oplus$5\%$\oplus$5\%    &   3\%$\oplus$5\%$\oplus$5\%              \\
$\ttbar$ background                &     10\%$\oplus$5\%   &   10\%$\oplus$5\%  &     10\%$\oplus$5\%   & 10\%$\oplus$5\%             \\
SM Higgs fact./renorm. scales      &     3.9 \%            &     3.9 \%         &     3.9 \%            &     3.9 \%            \\
SM Higgs PDF$+\alpS$            &     3.2 \%            &     3.2 \%         &     3.2 \%            &     3.2 \%         \\
[\cmsTabSkip]
Jet energy scale                   &        3--20\%        &        3--20\%     &          3--20\%       &   3--20\%                          \\
$\tauh$ energy scale              &        1.2\%          &        \NA       &           1.2\%       & \NA                               \\
$\PGm,\Pe\to\tauh$ energy scale   &           1.5\%       &       \NA        &           3\%         & \NA                                  \\
$\PGm$ energy scale                &        0.2\%          &        0.2\%       &          \NA          &  0.2\%                                 \\
$\Pe$ energy scale                 &       \NA           &         0.1--0.5\%  &      0.1--0.5\%        &  0.1--0.5\%                            \\
Unclustered energy scale           &        $\pm 1 \sigma$ &  $\pm 1 \sigma$    &      $\pm 1 \sigma$   &  $\pm 1 \sigma$                         \\
[\cmsTabSkip]
IntegRated luminosity              &              2.5\%    &       2.5\%        &          2.5\%        &   2.5\%                                         \\
\end{tabular}}
\end{table}

The uncertainties in the muon, electron and \PGt lepton  selection including the  trigger, identification (ID), and isolation efficiencies are estimated from collision data sets of $\PZ$ bosons decaying  to
$\Pe\Pe,\PGm\PGm,\PGt_{\PGm}\tauh$~\cite{Chatrchyan:2018xi,Khachatryan:2015hwa,Sirunyan:tau16}. They result in changes of normalisation, with the exception of the uncertainty on high \pt \PGt lepton efficiency that changes both yield and \mcol distribution shape.
The \PQb tagging efficiency  is measured in collision data, and the simulation is adjusted accordingly to match with it. The uncertainty in this measurement  is taken as the systematic error affecting the normalisation of the simulation~\cite{Sirunyan:2017ezt}.

The uncertainties in the estimate of the misidentified-lepton backgrounds ($\PGm\to\tauh$,
$\Pe\to\tauh$, $\text{jet}\to\tauh, \PGm, \Pe$) are extracted from the validation tests in
control samples, as described in Section~\ref{backgrounds}; they affect both the
normalisation and the shape of the \mcol distribution. The uncertainty in the QCD multijet
background yield is 30\%, and corresponds to the uncertainty in the extrapolation factor
from the same-sign to the opposite-sign region, as determined in Ref.~\cite{CMS-PAS-HIG-16-043}. The uncertainties in the background contributions from
$\PZ$, $\PW\PW, \PZ\PZ$, $\PW\gamma$, $\ttbar$ and single-top quark arise predominantly from those in the measured cross sections of these processes and are applied as uncertainties in sample normalisation.

The uncertainties in the Higgs boson production cross sections due to the factorisation and the renormalisation scales, as well as the PDFs and the strong coupling constant ($\alpS$),  result  in changes in normalisation. They are taken from Ref.~\cite{YR4} and summarized in Table~\ref{tab:systematics} for the SM Higgs boson and Table~\ref{HXS} for heavy Higgs bosons. Only effects on the total rate are considered. Effects on the acceptance have been neglected.

\begin{table}[!htbp]
  \topcaption{\label{HXS} Theoretical uncertainties from~\cite{YR4} are applied to the Higgs boson production cross sections for the different masses. In the reference, the PDF and $\alpS$ uncertainties are computed following the recommendation of the PDF4LHC working group. The remaining Gaussian uncertainty accounts for additional intrinsic sources of theory uncertainty described in detail in the reference.}

  \centering
  \begin{tabular} {cccc}
    $m_{\PH}$  (\GeVns) & Cross section (pb) &Theory, Gaussian (\%) & PDF$+\alpS$  (\%)\\\hline
200 & 16.94
&$\pm$1.8
&$\pm$3.0\\
300
&6.59
&$\pm$1.8
&$\pm$3.0\\
450
&2.30
&$\pm$2.0
&$\pm$3.1\\
600
&1
&$\pm$2.1
&$\pm$3.5\\
750
&0.50
&$\pm$2.1
&$\pm$4.0\\
900
& 0.27
&$\pm$2.2
&$\pm$4.6\\
\end{tabular}
\end{table}

Shape and normalization uncertainties arising from the uncertainty in the  jet energy scale are computed by propagating the effect of altering each source of jet energy scale uncertainty by $\pm 1$ standard deviation to the fit templates of each process. There are 27 independent sources of jet energy scale uncertainty, fully correlated between categories and $\PGt$ lepton decay channels.

The uncertainty in the $\tauh$ energy scale is treated equally for the two independent channels: $\PH \to \PGm \PGt_h$ and $\PH \to \Pe \PGt_h$. It is propagated to the collinear mass distributions. Also, the uncertainty in the energy scale of electrons and muons misidentified as \tauh is propagated to the \mcol distributions and produces changes in the distribution shape and normalization. Systematic uncertainties in the electron energy scale and resolution include the effects of electron selection efficiency,
pseudorapidity dependence and categorisation, summed in quadrature. They impact both the normalization and shape of the \mcol distribution. The systematic uncertainties in the energy resolution have negligible effect. The uncertainty in muon energy scale and resolution is also treated in the same manner.
Scale uncertainties on the energy from jets with \pt below 15\GeV and PF candidates not clustered inside jets  (unclustered energy scale uncertainty) are also considered \cite{Sirunyan:2019kia}. They  are estimated independently for four particle categories: charged particles, photons, neutral hadrons, and very forward  particles which are not contained in jets. The effect of shifting the energy of each particle by its uncertainty is propagated to \ptmiss and leads to  both changes in shape of the distribution and in overall predicted yields. The different systematic uncertainties from the four  particle categories, for the unclustered energy scale, are considered uncorrelated.

The bin-by-bin uncertainties \cite{BARLOW1993219} account for the statistical uncertainties in every bin of the template distributions of every process. They are uncorrelated between bins, processes, and categories.

Shape uncertainties related to the pileup have been considered  by varying  the weights applied to simulation.  This weight variation is obtained changing by 5\% the  total inelastic cross section used in the estimate of the pileup events in data~\cite{Sirunyan:2018nqx}. The new values are then applied, event by event, to produce alternative collinear mass distributions used as shape uncertainties in the fit. Other minimum bias event modelling and simulation uncertainties are estimated to be much smaller and are therefore neglected.  The uncertainty on the integrated luminosity affects all processes with normalization taken directly from simulation.

\section{Results\label{results}}

After all selection criteria have been applied, a binned maximum likelihood fit is performed on the $\mcol$  distributions in the range 0--1400\GeV, looking for a signal-like excess on top of the estimated  background. No fit on the control region is performed.  The systematic uncertainties are represented by nuisance parameters, assuming a log normal prior for normalisation parameters, and Gaussian priors for $\mcol$ shape uncertainties. The uncertainties that affect the shape of the $\mcol$ distribution, mainly those corresponding to the energy scales, are represented by nuisance parameters whose variation results in a modification of the distribution~\cite{Conway:2011in, LHC-HCG-Report}. A profile likelihood ratio is used as test statistic. The $95$\% \CL upper limits on the \PH production cross section times branching fraction to LFV lepton channels, $\sigma(\Pg\Pg\to \PH)\mathcal{B}(\PH\to\PGm\PGt)$ and $\sigma(\Pg\Pg\to \PH)\mathcal{B}(\PH\to\Pe\PGt)$,  are set using the \CLs criterion~\cite{Junk,Read2} and the asymptotic approximation of the distributions of the LHC test-statistic~\cite{CLs}, in a combined fit to the $\mcol$ distributions. The limits are also computed per channel and category. The upper limits are derived in the analysed mass range in steps of 50\GeV. Where simulated samples are not available, shapes and yields are interpolated from the neighbouring samples with a moment morphing algorithm that derive the $\mcol$ distribution from the two closest simulated mass points.

\subsection{\texorpdfstring{\Hmt}{H to mu tau} results}
\begin{figure}[htbp]
  \centering
     \includegraphics[width=0.49\textwidth]{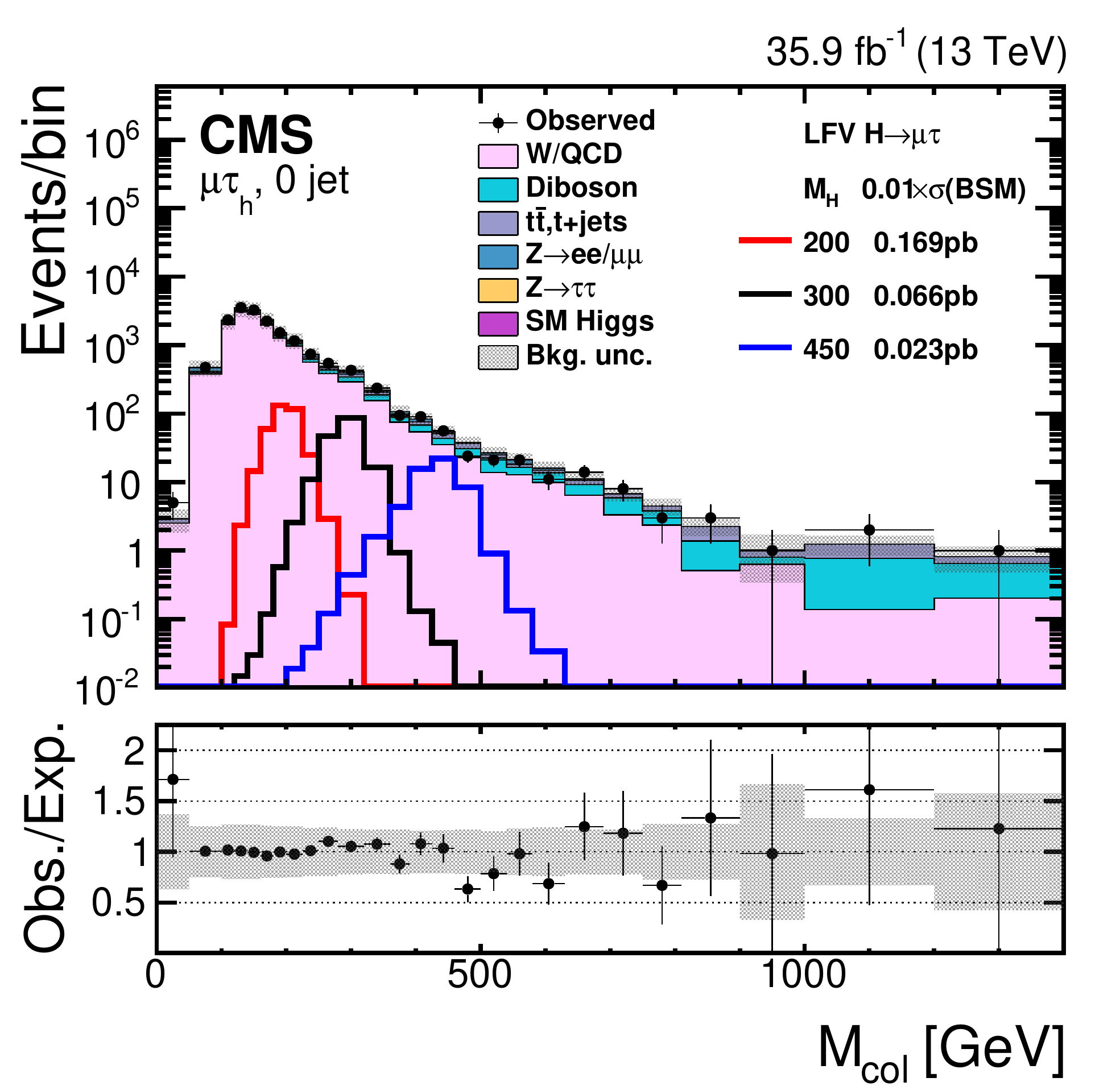}
     \includegraphics[width=0.49\textwidth]{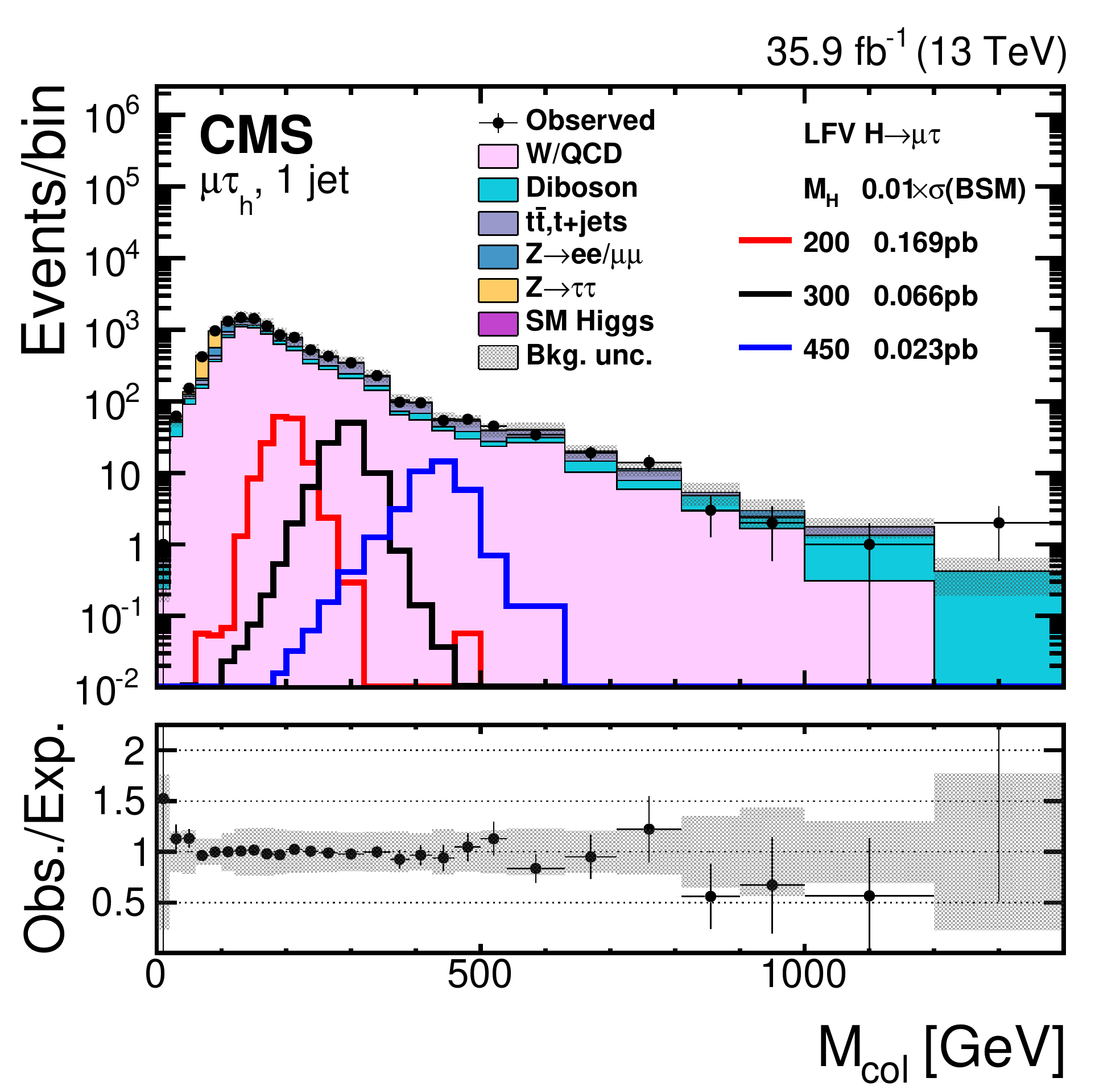}\\
     \includegraphics[width=0.49\textwidth]{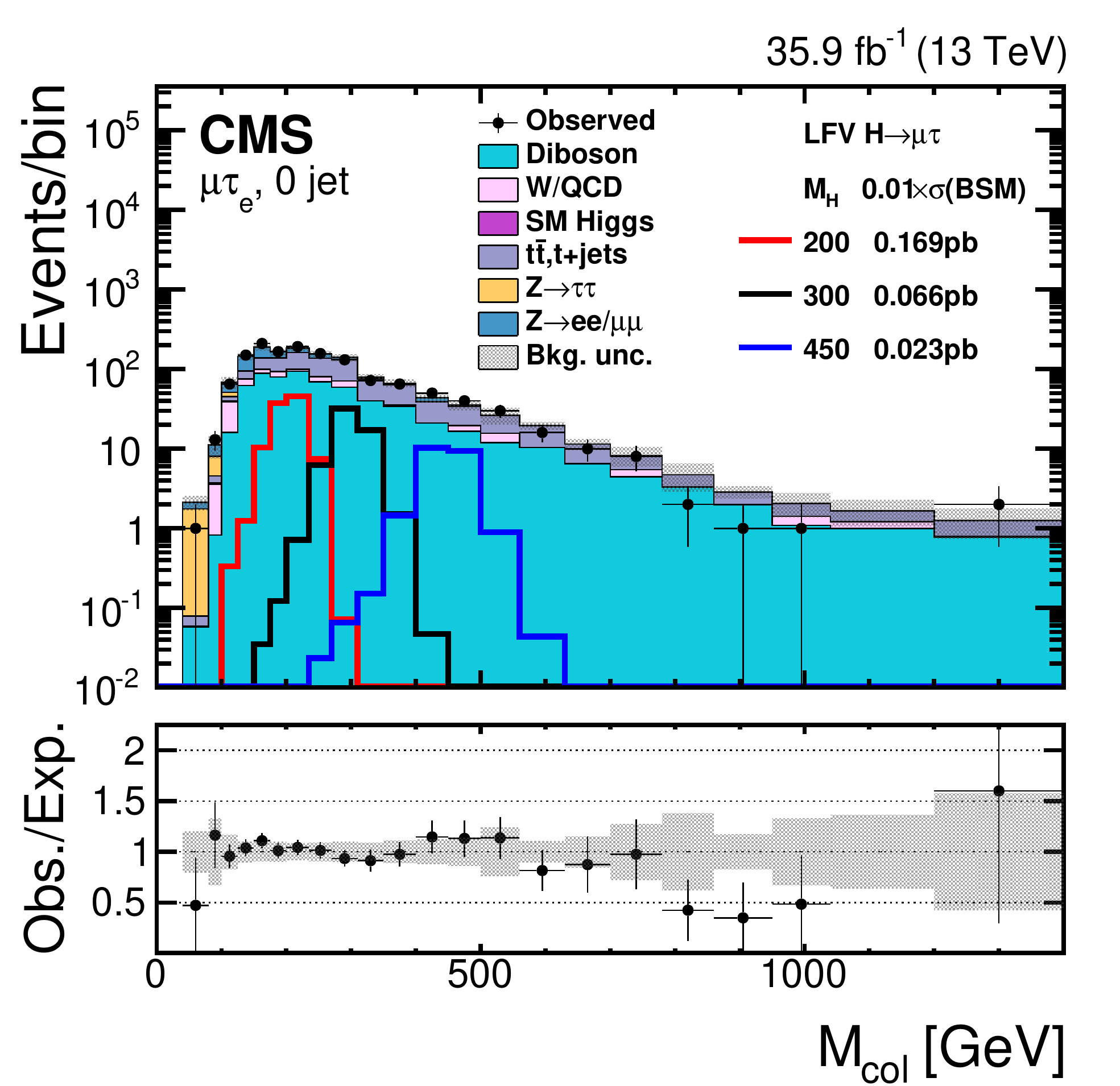}
     \includegraphics[width=0.49\textwidth]{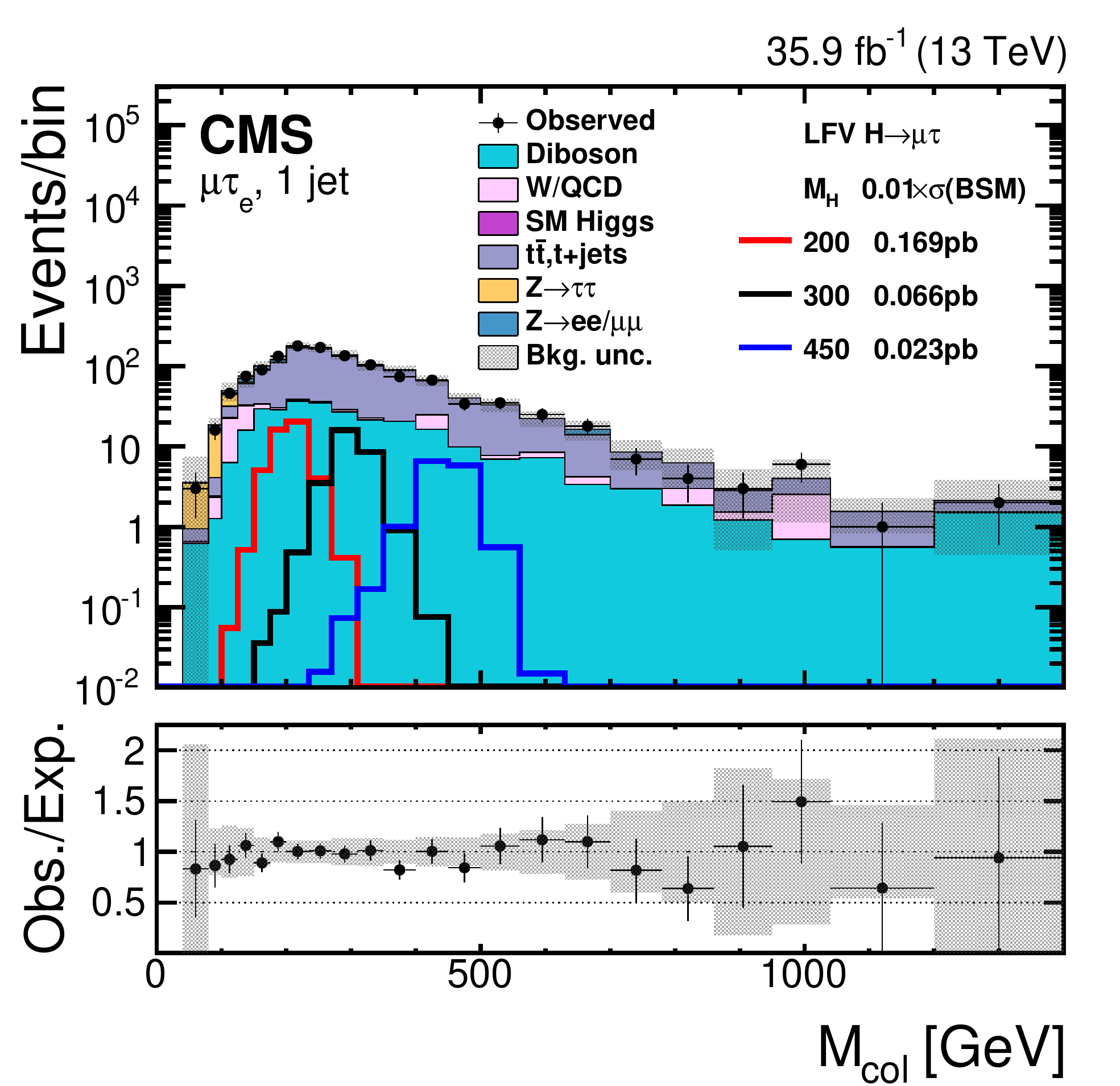}
     \caption{The \mcol distribution in the signal region, for the $\PGm\tauh$ (upper) and $\PGm\PGt_\Pe$ (lower) channels for the Higgs boson mass in the range 200--450\GeV for 0-jet (left) and 1-jet (right) categories. The uncertainty bands include both statistical and systematic uncertainties. The plotted values are number of events per bin using a variable bin size. The background is normalised to the best fit values from a binned likelihood fit, discussed in the text, to the background only hypothesis. For depicting the signals a branching fraction of 1\% and BSM cross sections from Ref.~\cite{YR4} are assumed.
     \label{masslowMassMuTau}}

\end{figure}

\begin{figure}[htbp]
  \centering
     \includegraphics[width=0.49\textwidth]{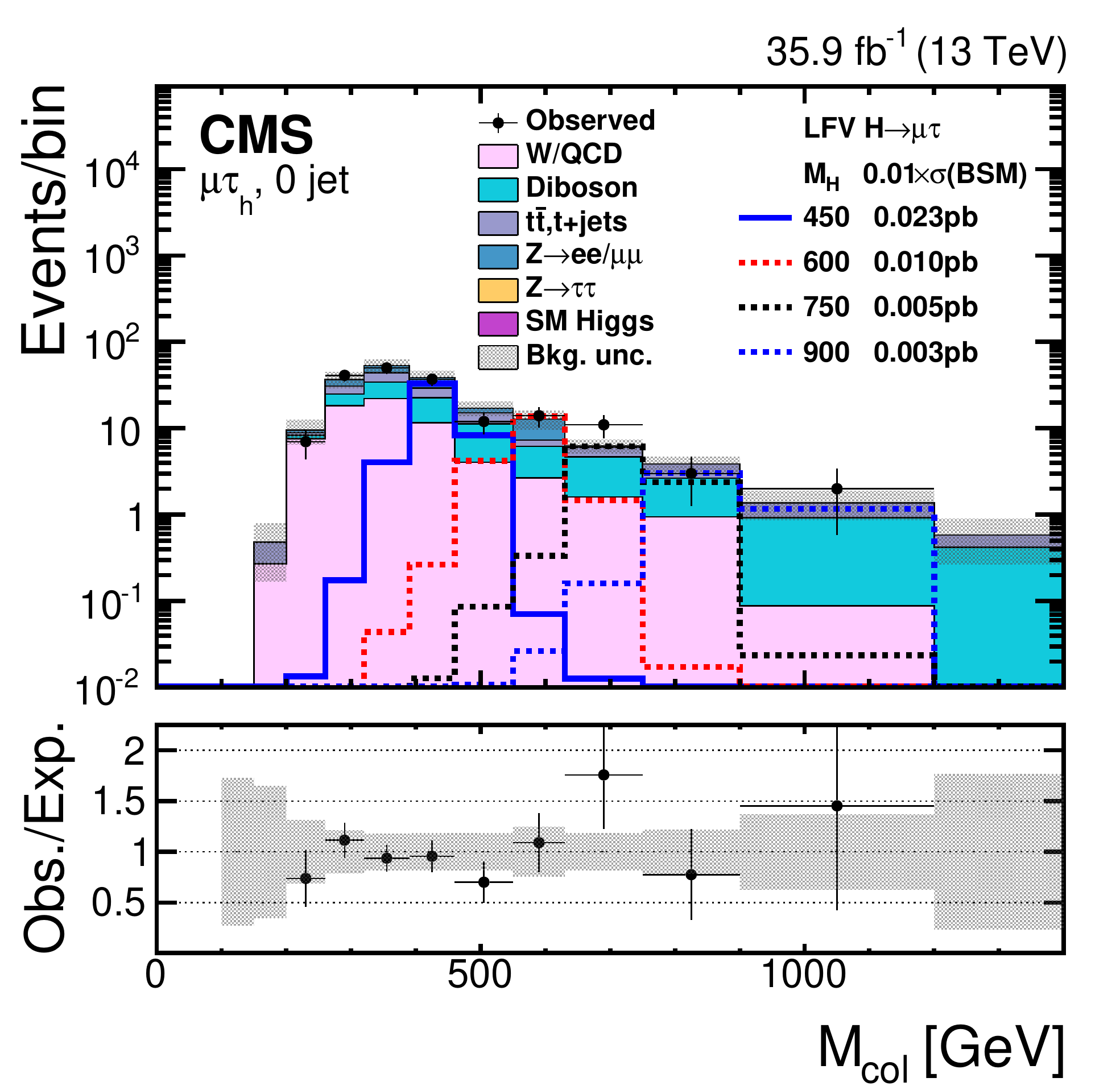}
     \includegraphics[width=0.49\textwidth]{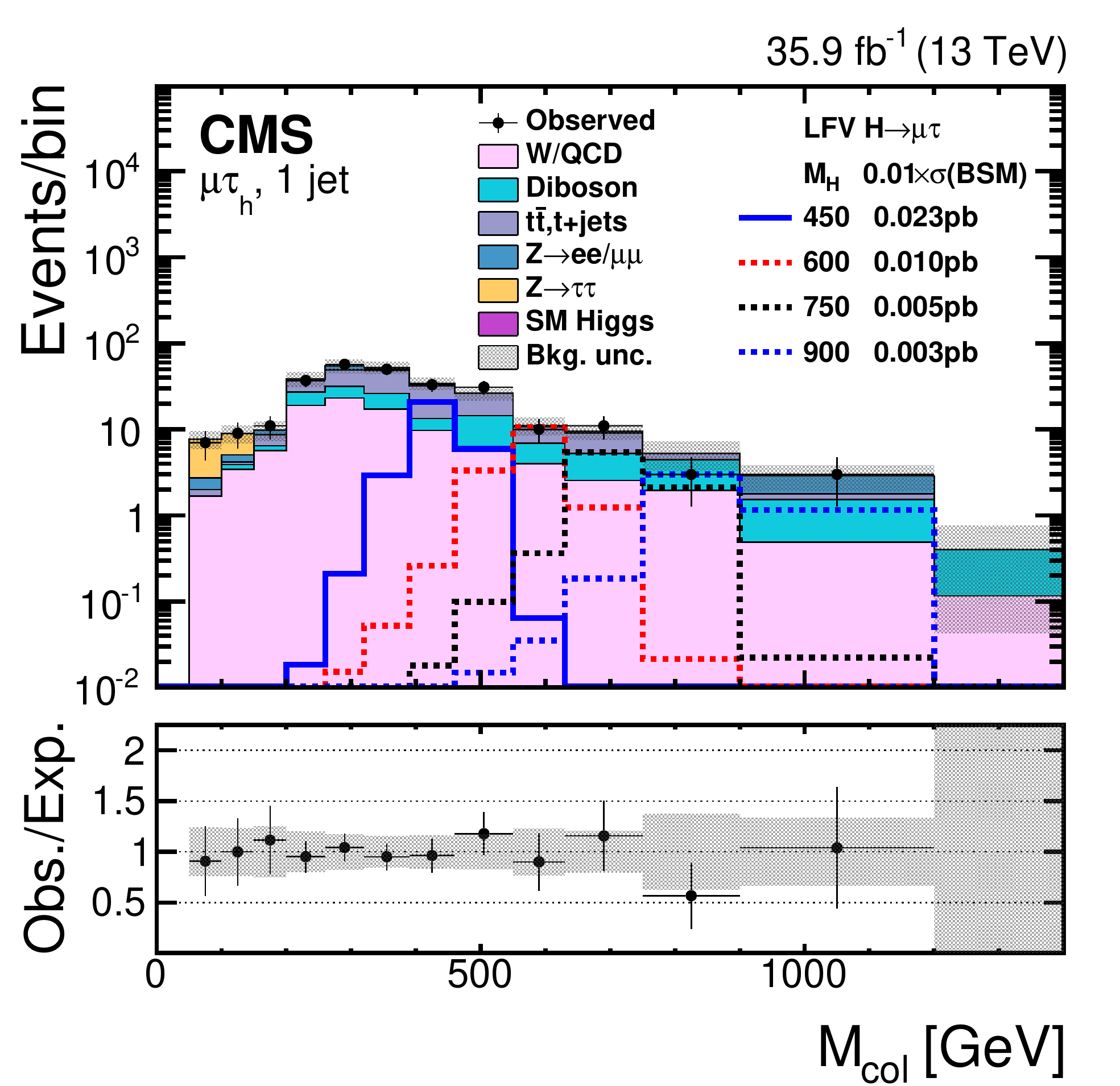}\\
     \includegraphics[width=0.49\textwidth]{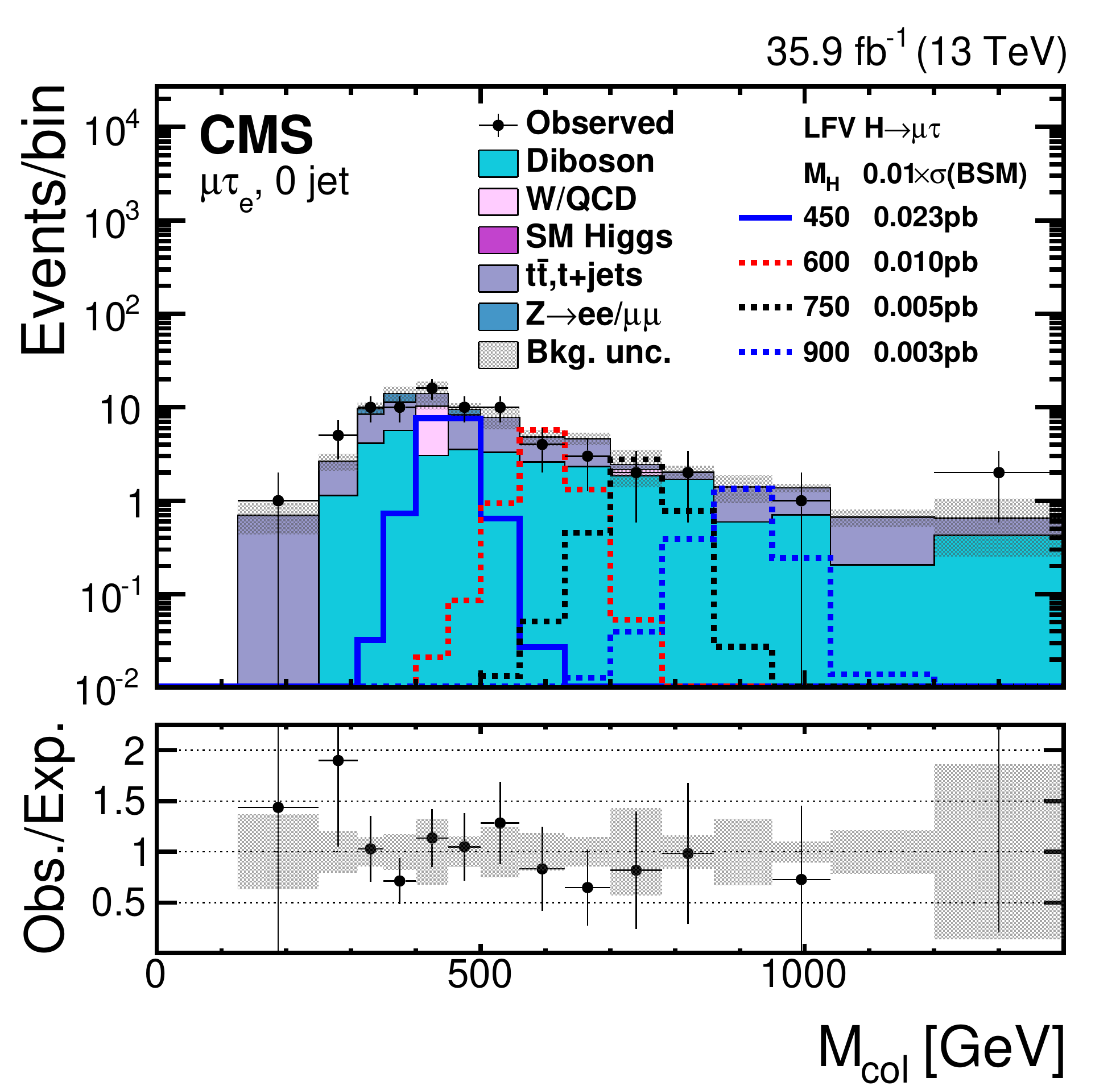}
     \includegraphics[width=0.49\textwidth]{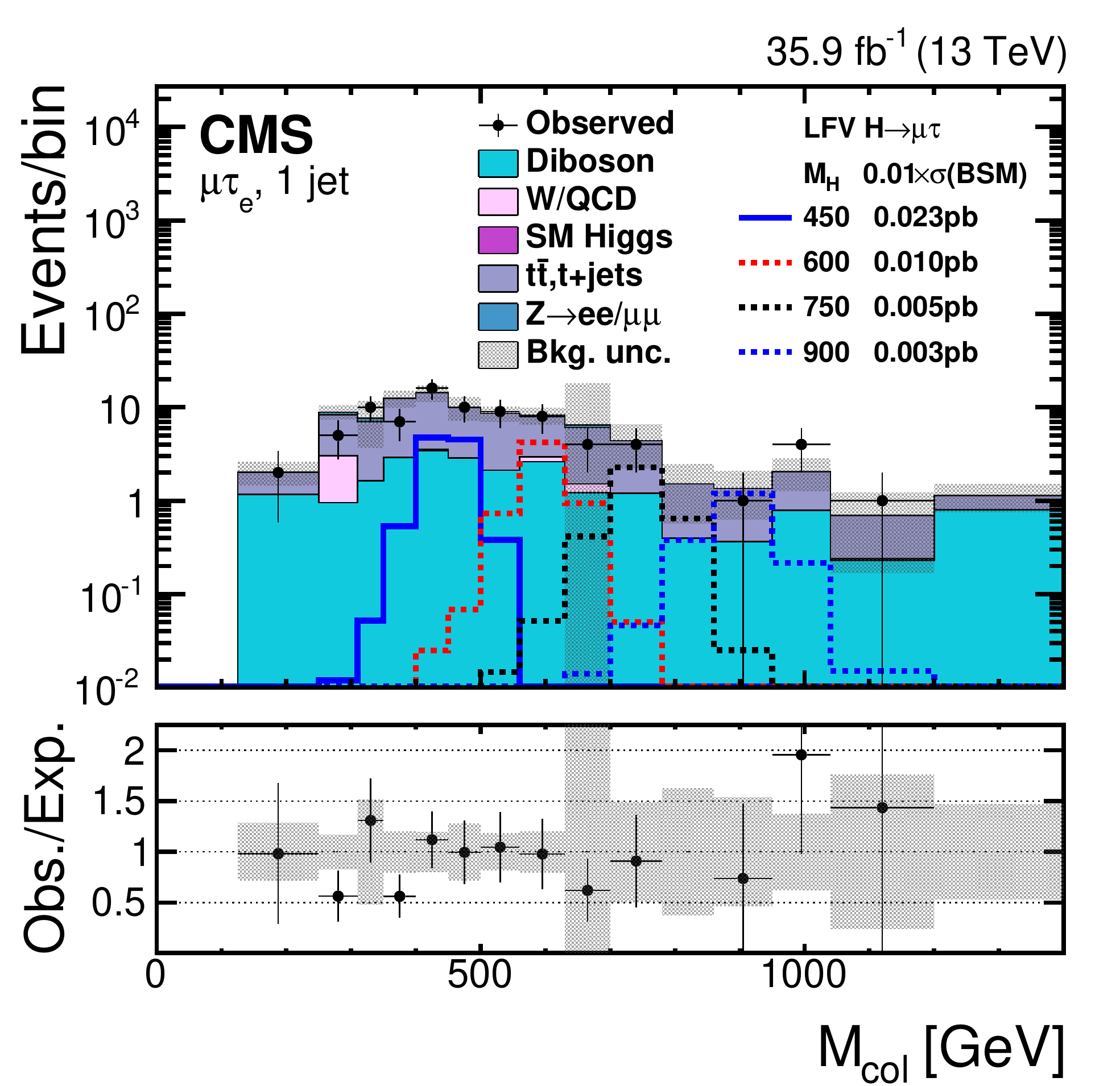}
     \caption{The \mcol distribution in the signal region, for the $\PGm\tauh$ (upper) and $\PGm\PGt_\Pe$ (lower) channels for the Higgs boson mass in the range 450--900\GeV for 0-jet (left) and 1-jet (right) categories. The uncertainty bands include both statistical and systematic uncertainties. The plotted values are number of events per bin using a variable bin size. The background is normalised to the best fit values from a binned likelihood fit, discussed in the text, to the background only hypothesis. For depicting the signals a branching fraction of 1\% and BSM cross sections from Ref.~\cite{YR4} are assumed.
     \label{masshighMassMuTau}}

\end{figure}

\begin{figure}[htbp]
  \centering
     \includegraphics[width=0.49\textwidth]{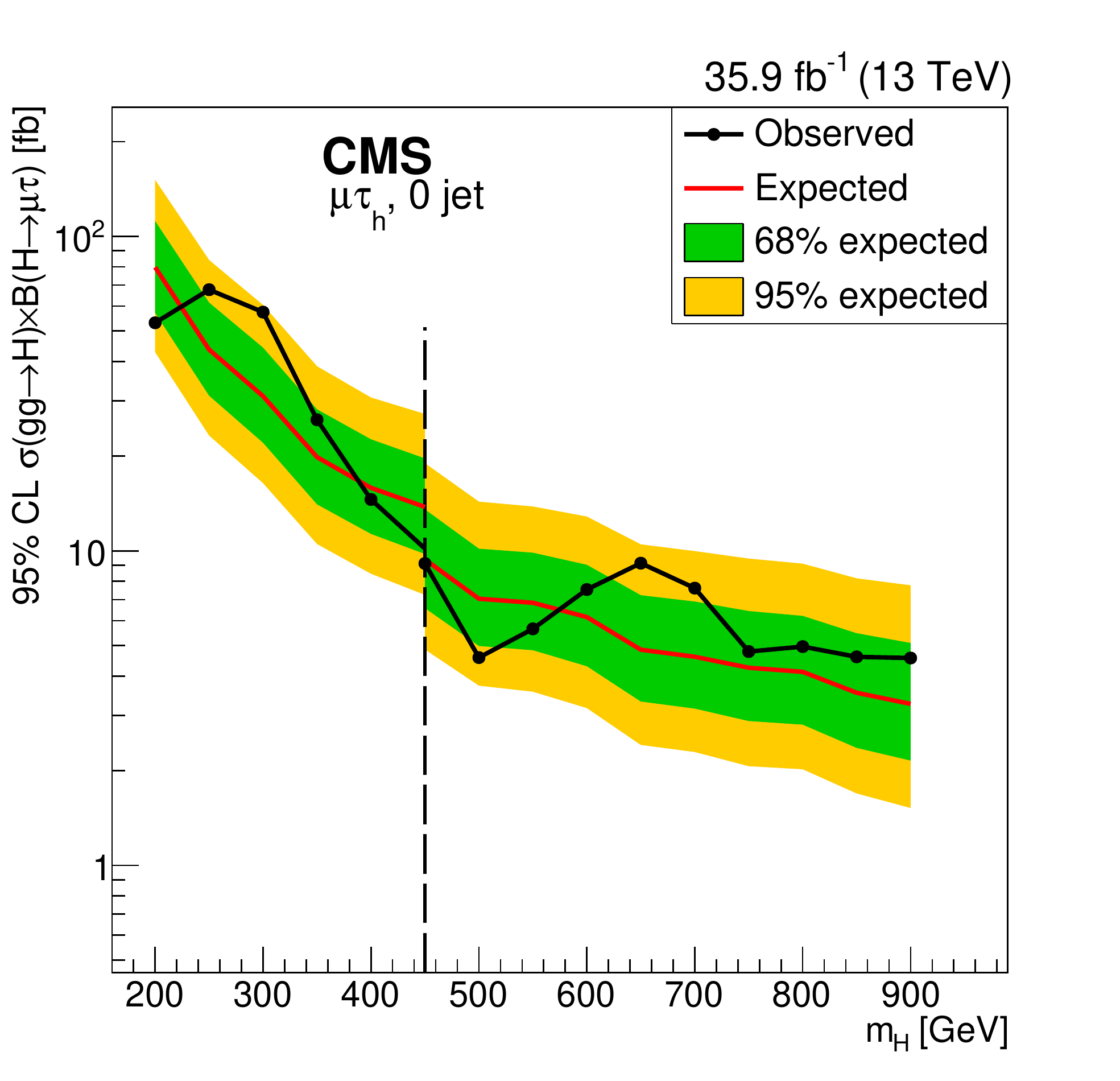}
     \includegraphics[width=0.49\textwidth]{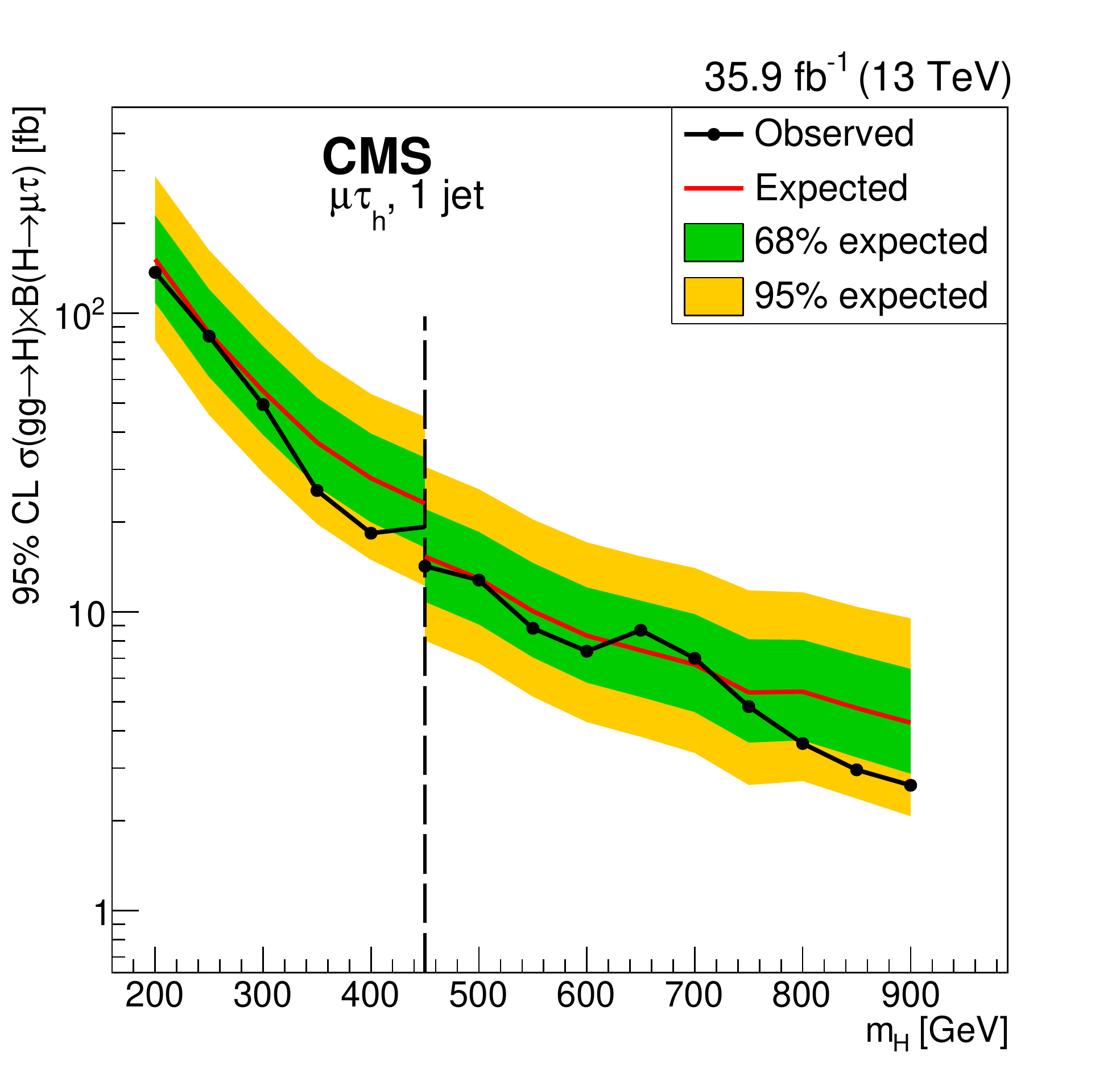}\\
     \includegraphics[width=0.49\textwidth]{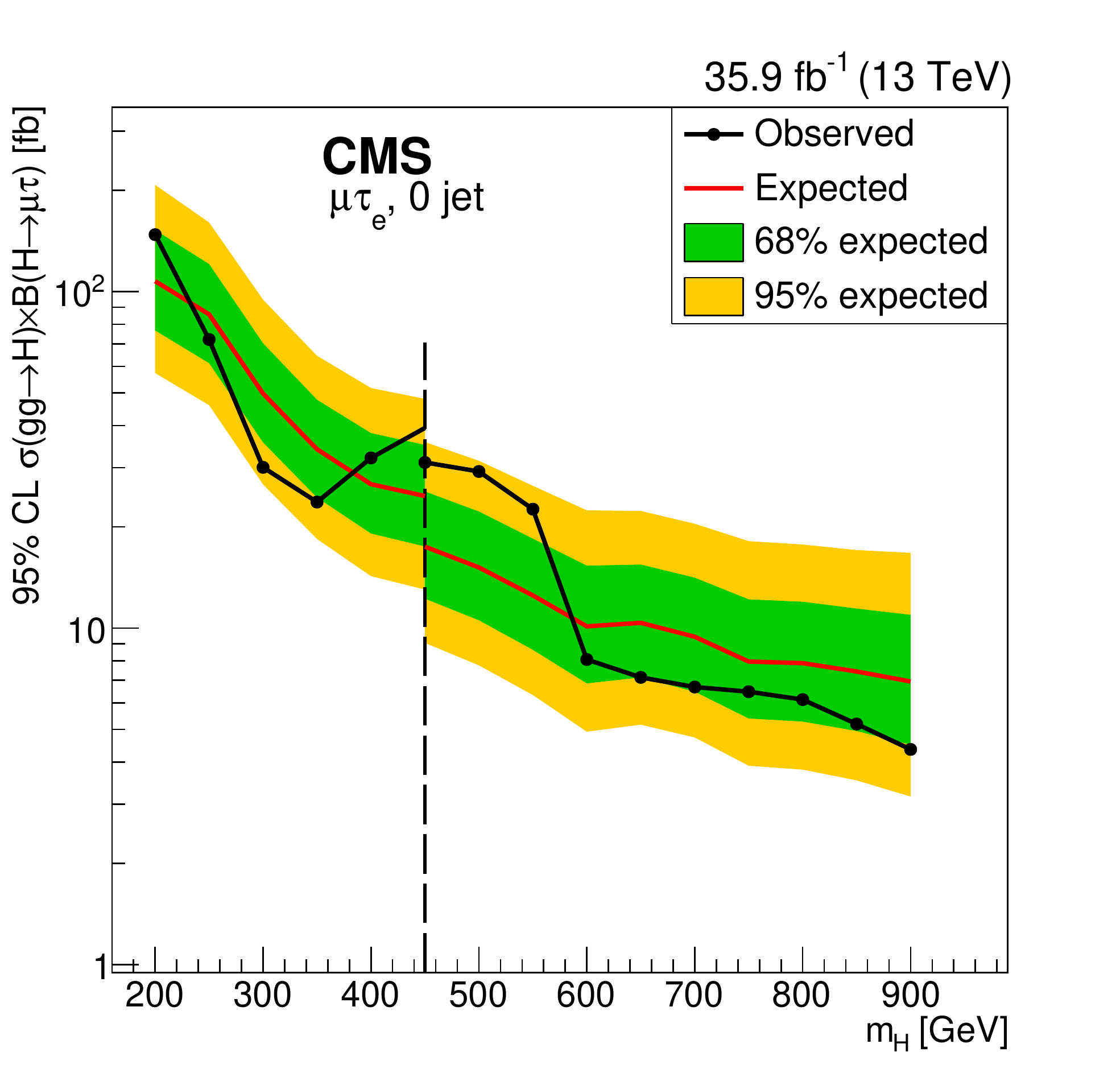}
     \includegraphics[width=0.49\textwidth]{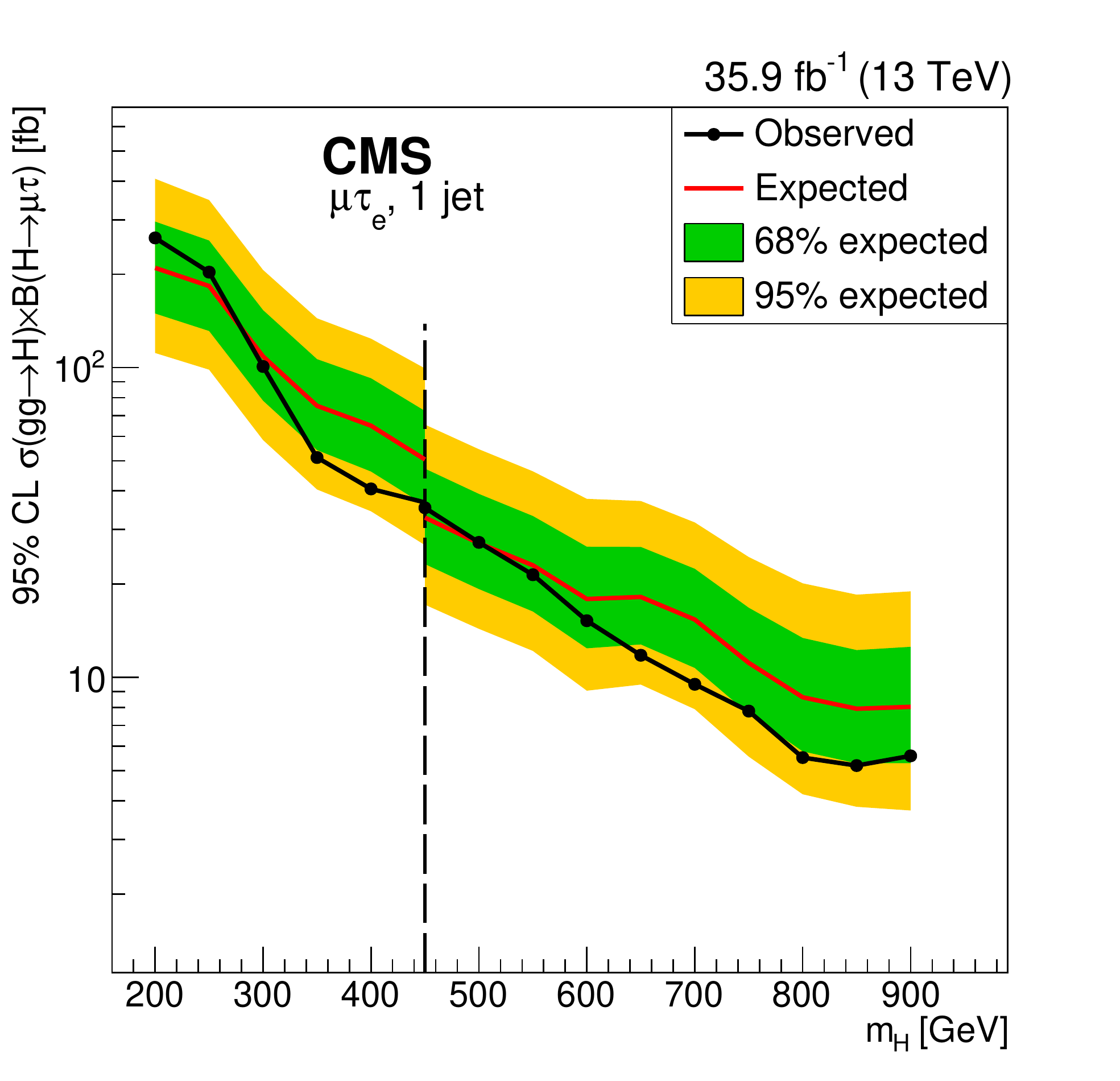}
     \caption{The observed and median expected 95\% \CL upper limits on $\sigma(\Pg\Pg\to \PH)\mathcal{B}(\PH\to\PGm\PGt)$, for the $\PGm\tauh$ (upper) and $\PGm\PGt_\Pe$ (lower) channels, for 0-jet (left) and 1-jet (right) categories.  The dashed line shows the transition between the two investigated mass ranges.}
     \label{fig:masslimitsMUTAUcategories}
\end{figure}

\begin{figure}[htbp]
     \centering
     \includegraphics[width=0.49\textwidth]{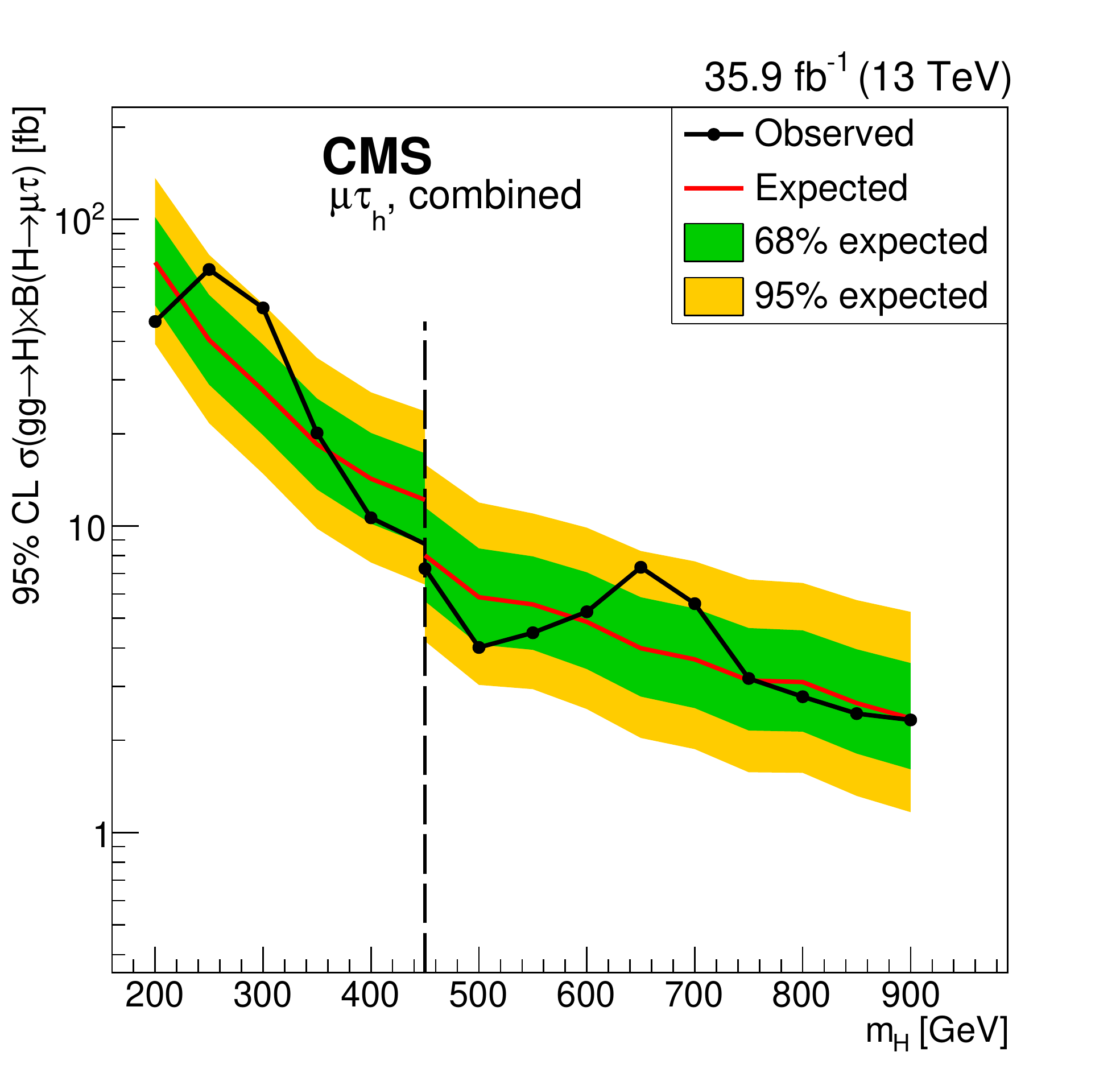}
     \includegraphics[width=0.49\textwidth]{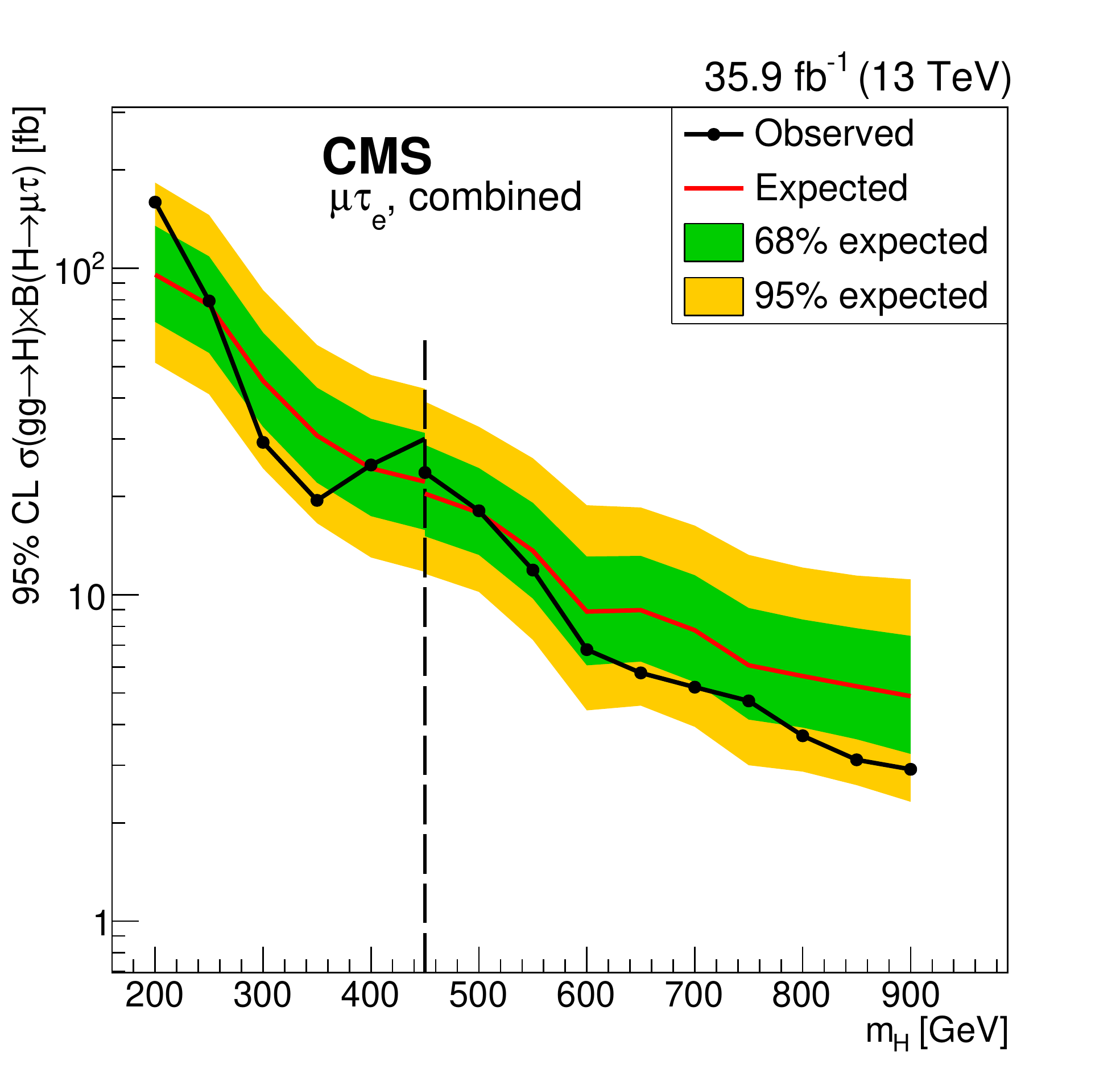}\\
     \includegraphics[width=0.8\textwidth ]{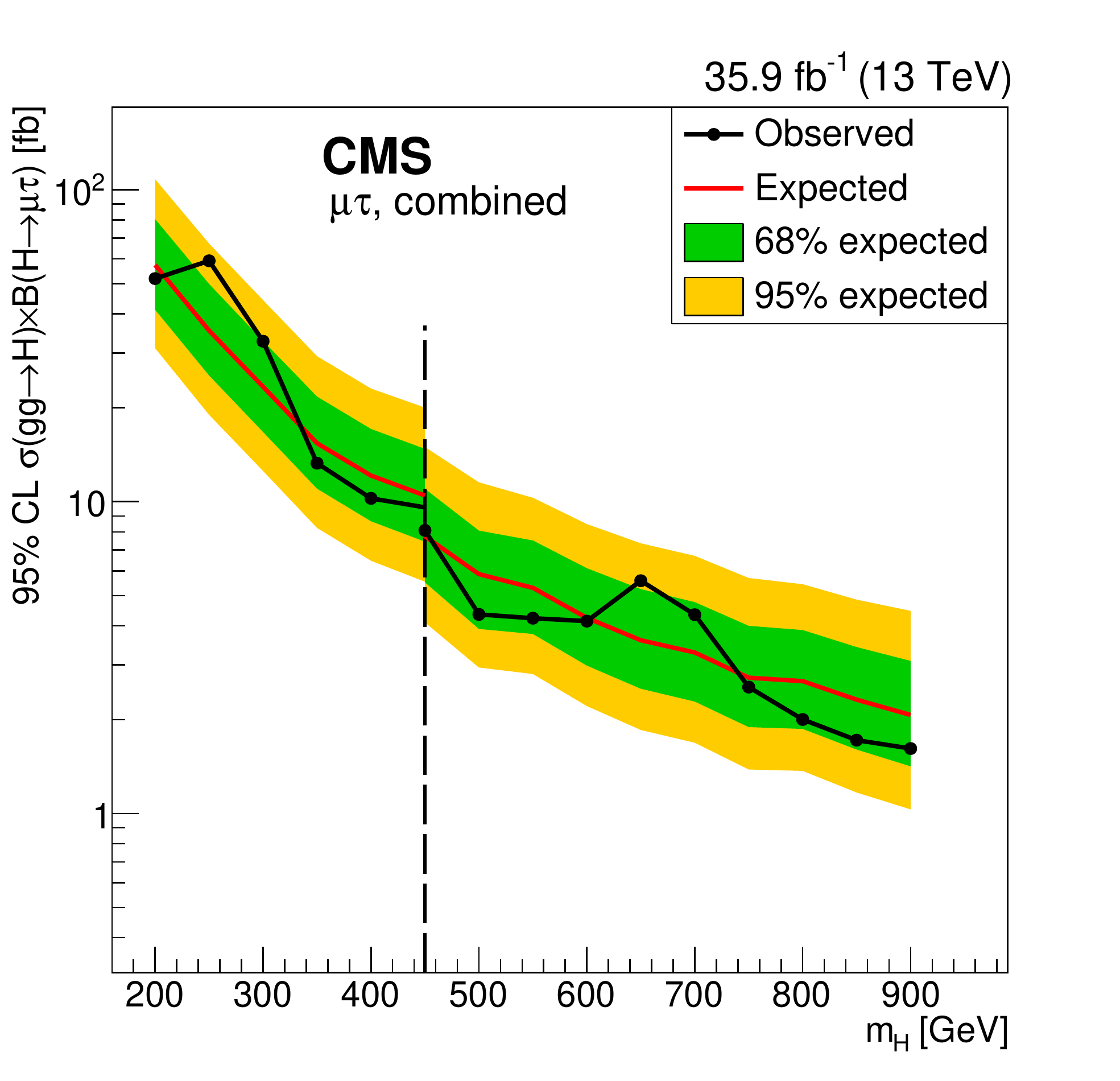}
     \caption{The combined observed and median expected 95\% \CL upper limits on $\sigma(\Pg\Pg\to \PH)\mathcal{B}(\PH\to\PGm\PGt)$, for $\PGm\tauh$ (upper left) and $\PGm\PGt_\Pe$ (lower right) channels, and their combination $\PGm\PGt$ (lower).  The dashed line shows the transition between the two investigated mass ranges.}
     \label{fig:masslimitsMUTAUcombined}
\end{figure}

The distributions of the collinear mass $\mcol$ compared to the signal and background contributions in the $\Hmuhad$ and $\Hmue$ channels, in each jet category, are shown in Figs.~\ref{masslowMassMuTau} and~\ref{masshighMassMuTau}. No excess over the background expectation is observed.
The observed and median expected 95\% \CL upper limits on $\sigma(\Pg\Pg\to \PH)\mathcal{B}(\PH \to \PGm \PGt)$ range from 51.9 (57.4)\unit{fb} to 1.6 (2.1)\unit{fb}, and are given for each category in Table~\ref{tab:MUTAULIMITSTABLE}.The limits are also summarized graphically in Fig.~\ref{fig:masslimitsMUTAUcategories} for the individual categories, and in Fig.~\ref{fig:masslimitsMUTAUcombined} for the combination of the two $\PGt$ decay channels.

\begin{table}
\topcaption{The observed and median expected 95\% \CL upper limits on $\sigma(\Pg\Pg\to \PH) \mathcal{B}(\PH\to\PGm\PGt)$.}\label{tab:MUTAULIMITSTABLE}
\centering
\begin{tabular}{cccccccccccc}
\multicolumn{12}{c}{Observed 95\% \CL upper limit on $\sigma(\Pg\Pg\to \PH)\mathcal{B}(\PH\to\PGm\PGt)$ (fb)} \\
\hline
& \multicolumn{3}{c}{$\PGm\PGt_\Pe$} && \multicolumn{3}{c}{$\PGm\tauh$} &&\multicolumn{3}{c}{$\PGm\PGt$}  \\
$m_{\PH} $ (\GeVns) & 0 jet & 1 jet  & comb && 0 jet & 1 jet  & comb && 0 jet & 1 jet  & comb \\
\hline
200 &147.8& 262.1& 159.4 &&  53.1 & 136.9& 46.4&&53.3 &133.9 & 51.9\\
300 &30.1& 100.8& 29.3 &&  57.4 & 49.4& 51.4&&33.2 &45.5 & 32.7\\
450 &31.1& 35.3& 23.7 &&  9.1 & 14.2& 7.3&&14.7 &14.6 & 8.1\\
600 &8.1& 15.2& 6.8 &&  7.5 & 7.4& 5.3&&9.1 &6.5 & 4.1\\
750 &6.5& 7.8& 4.7 &&  4.8 & 4.8& 3.2&&3.6 &3.7 & 2.5\\
900 &4.4& 5.6& 2.9 &&  4.6 & 2.6& 2.3&&3.0 &2.1 & 1.6\\
\end{tabular}
\\[\cmsTabSkip]
\centering
\begin{tabular}{cccccccccccc}
\multicolumn{12}{c}{Median expected 95\% \CL upper limit on $\sigma(\Pg\Pg\to \PH)\mathcal{B}(\PH\to\PGm\PGt)$ (fb) } \\
\hline
& \multicolumn{3}{c}{$\PGm\PGt_\Pe$} && \multicolumn{3}{c}{$\PGm\tauh$} &&\multicolumn{3}{c}{$\PGm\PGt$}  \\
$m_{\PH}$ (\GeVns) & 0 jet & 1 jet  & comb && 0 jet & 1 jet  & comb && 0 jet & 1 jet  & comb \\
\hline
200 &107.5& 209.8& 95.6 &&  79.7 & 151.6& 72.5&&63.7 &126.1 & 57.4\\
300 &49.8& 108.6& 45.2 &&  31.0 & 54.8& 27.7&&25.9 &48.8 & 23.4\\
450 &17.5& 32.8& 20.4 &&  9.4 & 15.3& 8.0&&8.2 &13.6 & 7.7\\
600 &10.4& 17.9& 8.9 &&  6.2 & 8.3& 4.9&&5.1 &7.4 & 4.2\\
750 &8.0& 11.1& 6.1 && 4.3 & 5.4& 3.1&&3.6 &4.7 & 2.7\\
900 &6.9& 8.0& 4.9 &&  3.3 & 4.3& 2.4&&2.8 &3.5 & 2.1\\
\end{tabular}
\end{table}

\subsection{\texorpdfstring{\Het}{H to e tau} results}

\begin{figure}[htbp]
  \centering
       \includegraphics[width=0.49\textwidth]{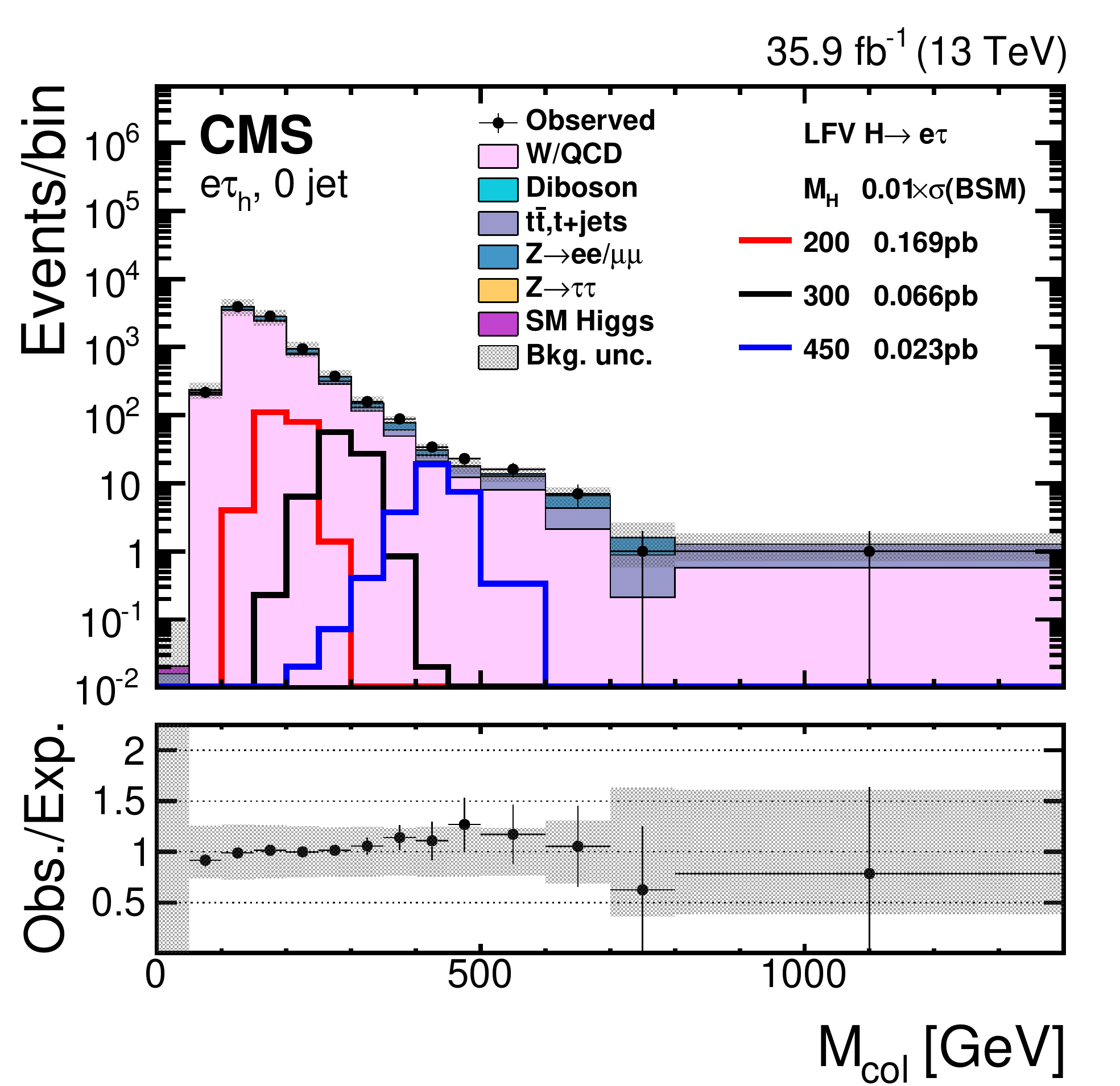}
     \includegraphics[width=0.49\textwidth]{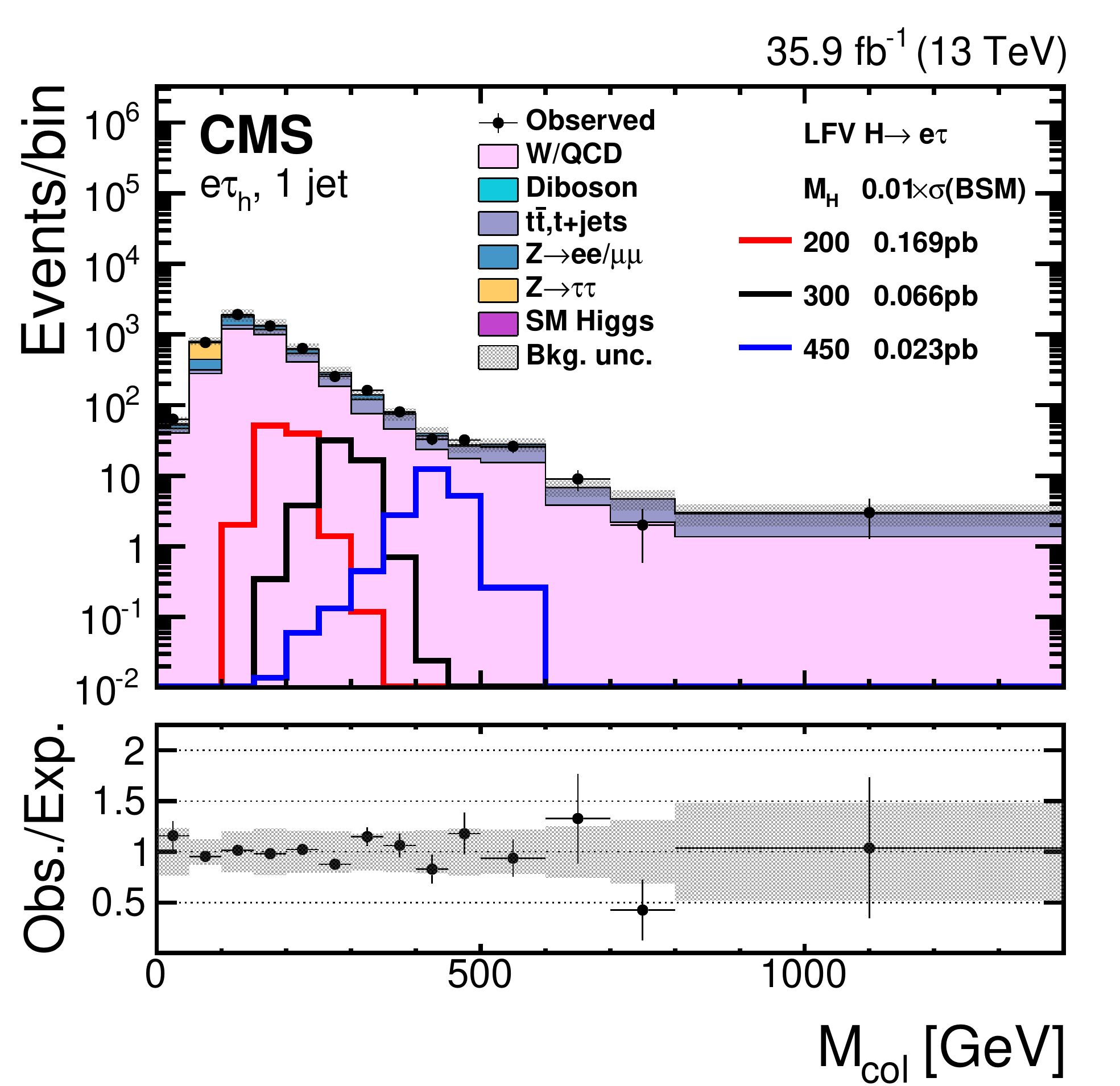}\\
     \includegraphics[width=0.49\textwidth]{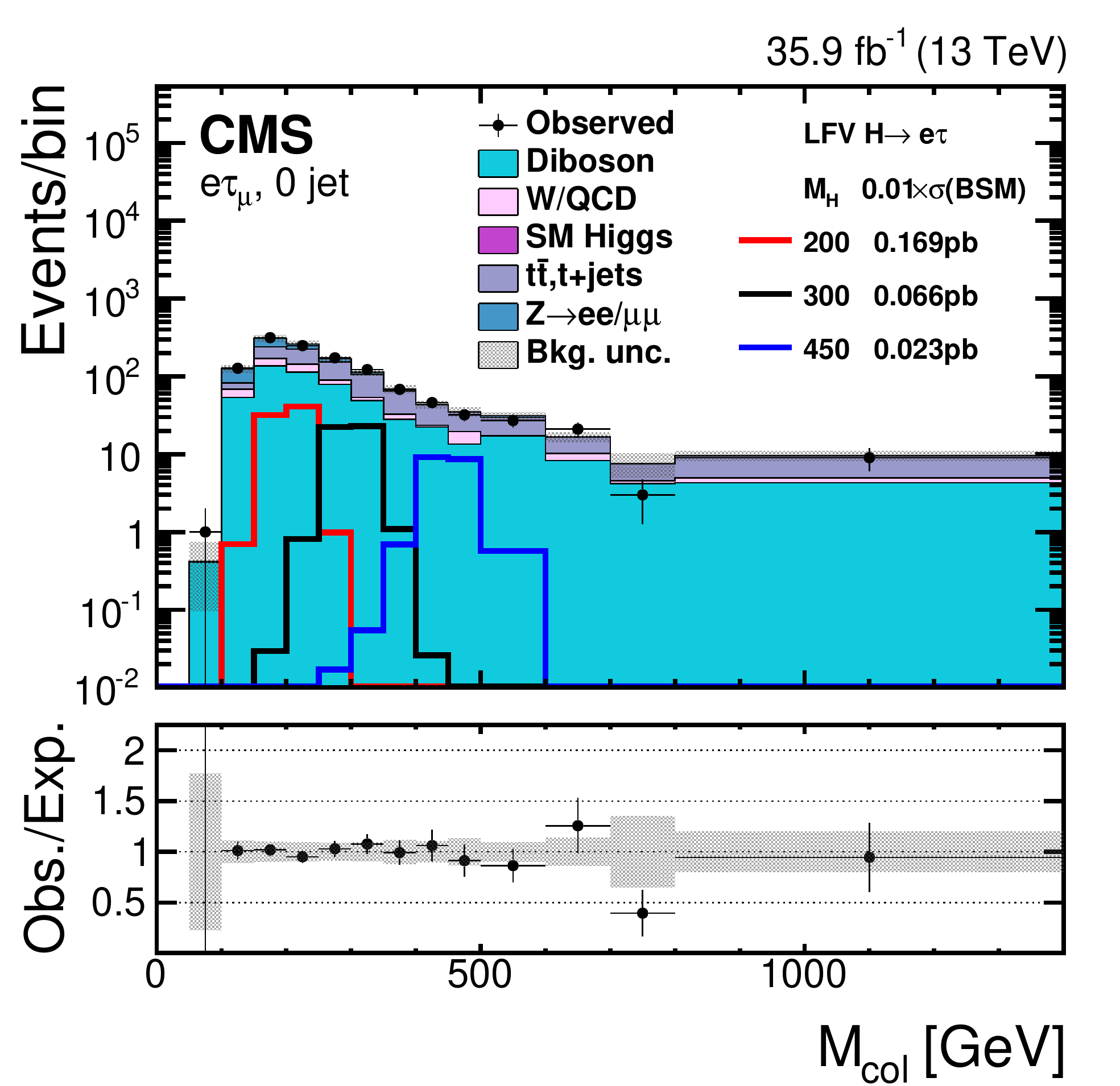}
     \includegraphics[width=0.49\textwidth]{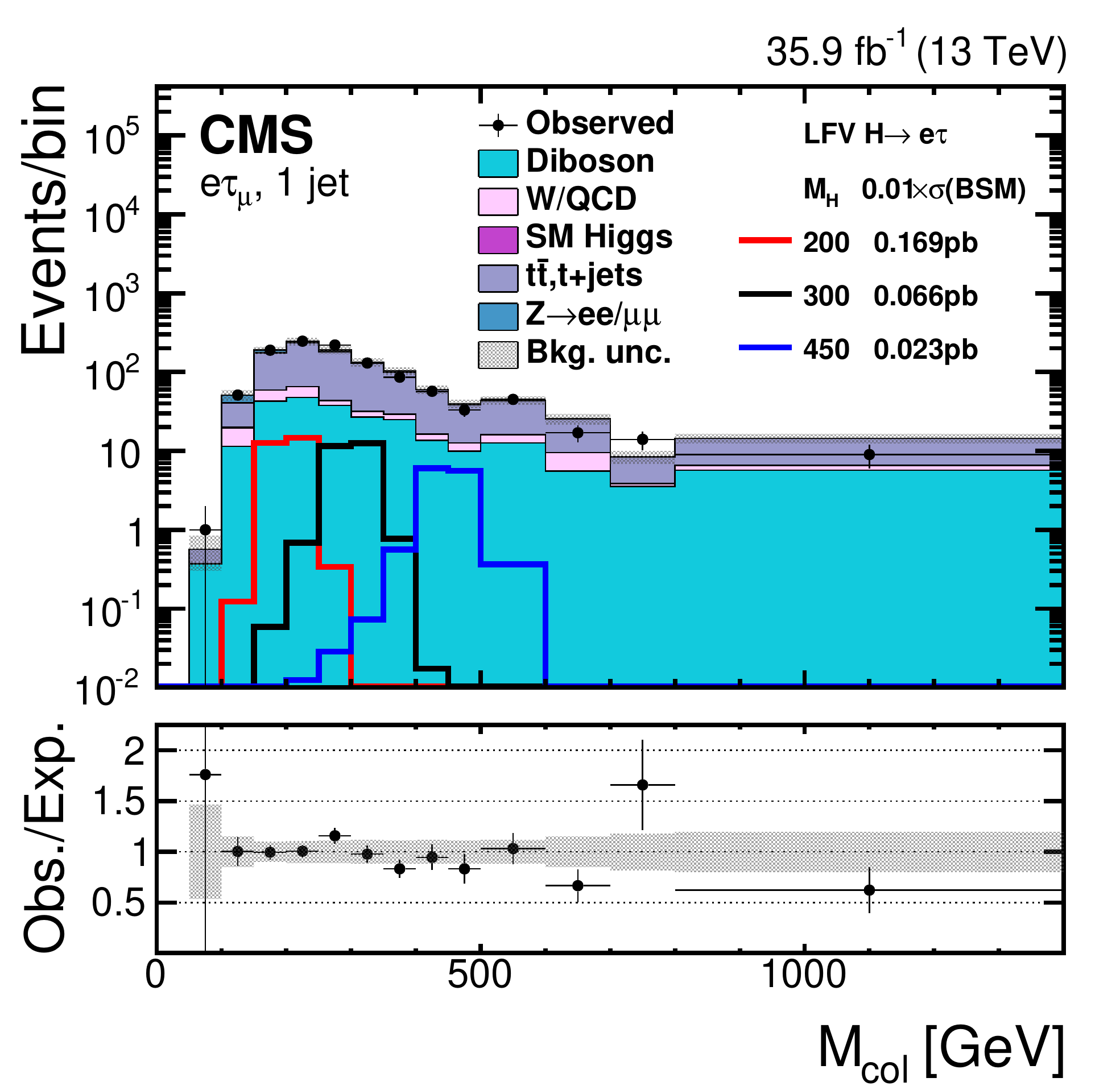}
     \caption{The \mcol distribution in the signal region, for the $\Pe\tauh$ (upper) and $\Pe\PGt_\PGm$ (lower) channels for the Higgs boson mass in the range 200--450\GeV for 0-jet (left) and 1-jet (right) categories. The uncertainty bands include both statistical and systematic uncertainties. The plotted values are number of events per bin using a variable bin size. The background is normalised to the best fit values from a binned likelihood fit, discussed in the text, to the background only hypothesis. For depicting the signals a branching fraction of 1\% and BSM cross sections from Ref.~\cite{YR4} are assumed.}
     \label{fig:McolSignalRegionETau}
\end{figure}

\begin{figure}[htbp]
  \centering
     \includegraphics[width=0.49\textwidth]{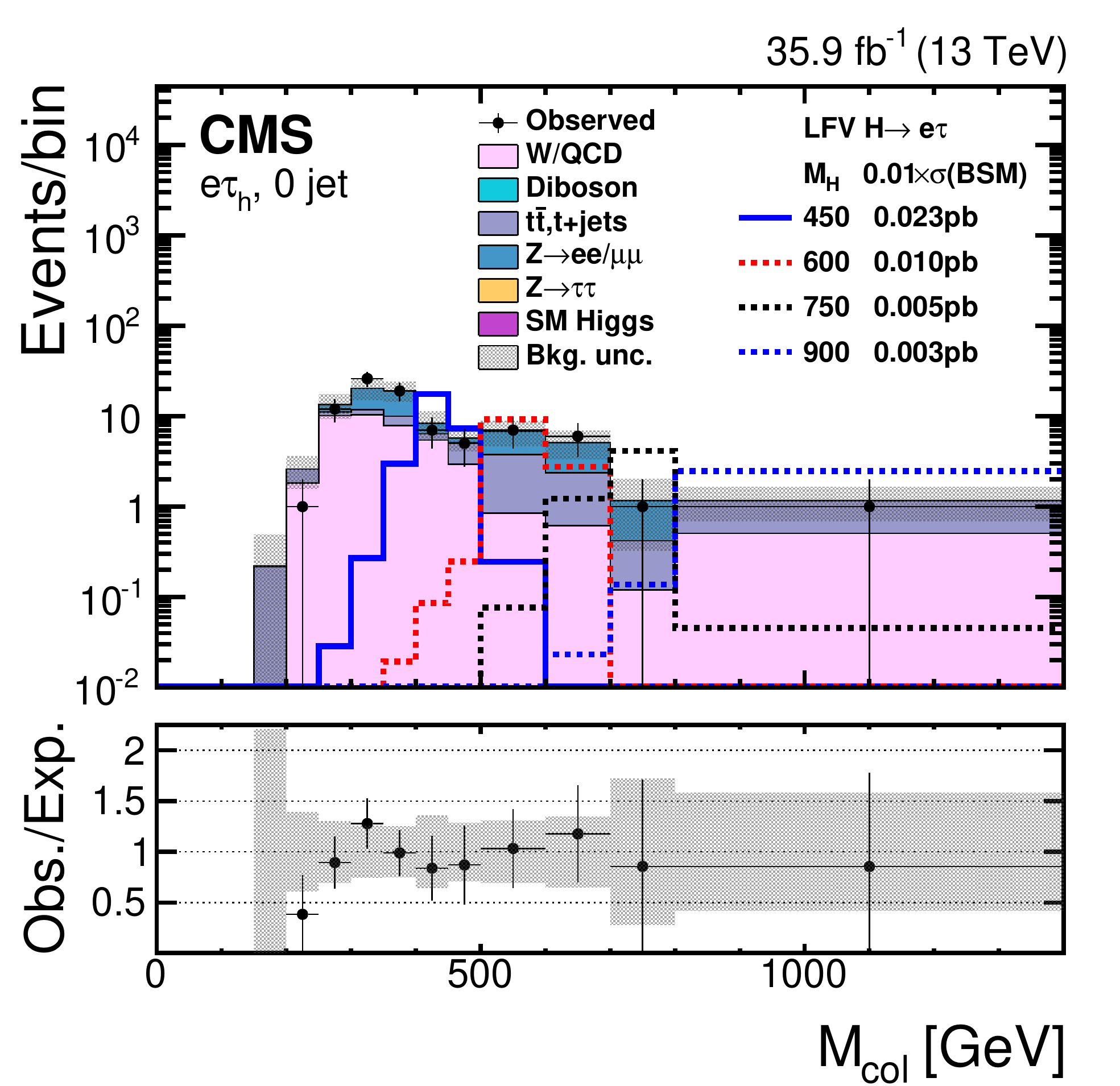}
     \includegraphics[width=0.49\textwidth]{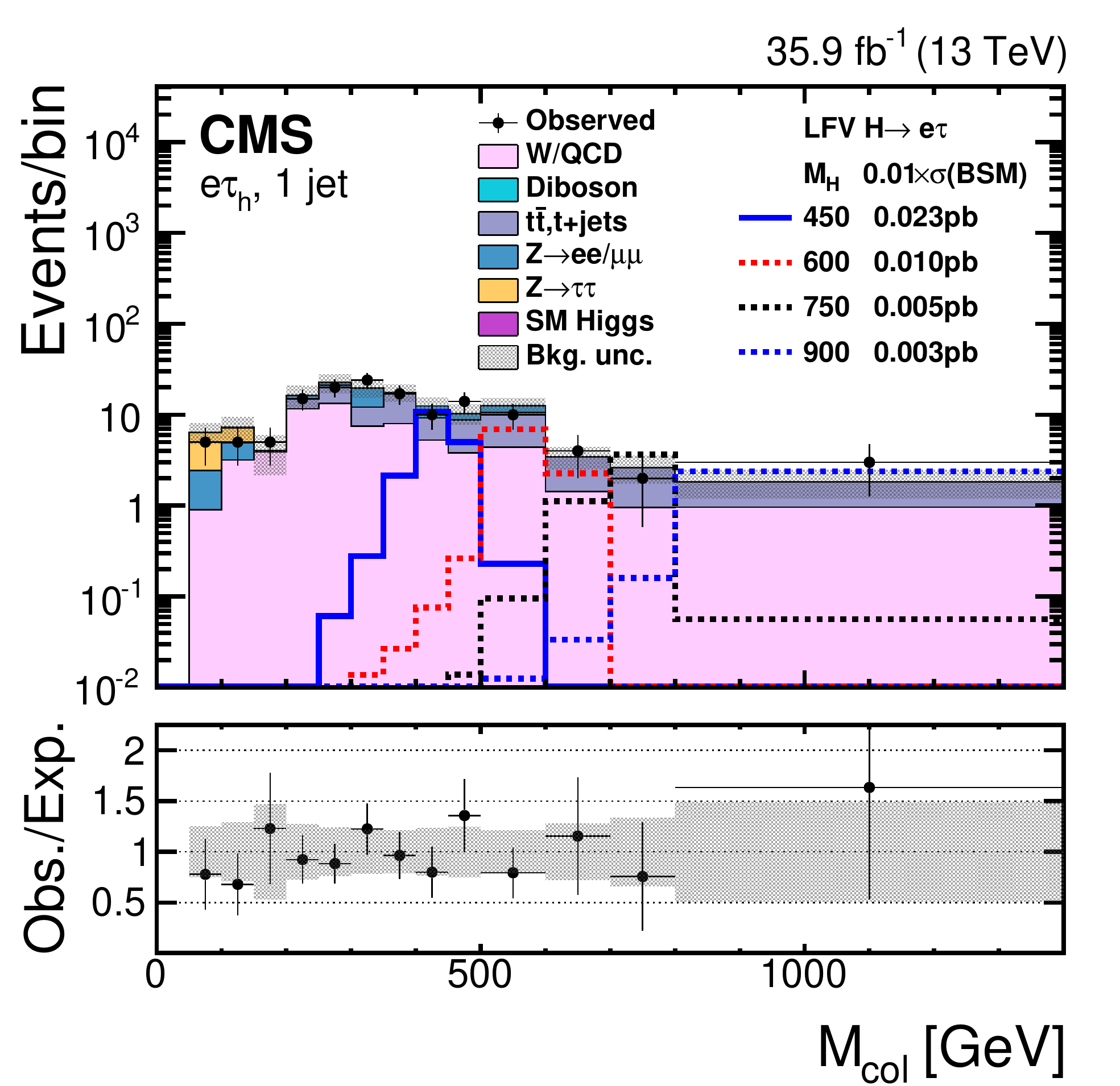}\\
     \includegraphics[width=0.49\textwidth]{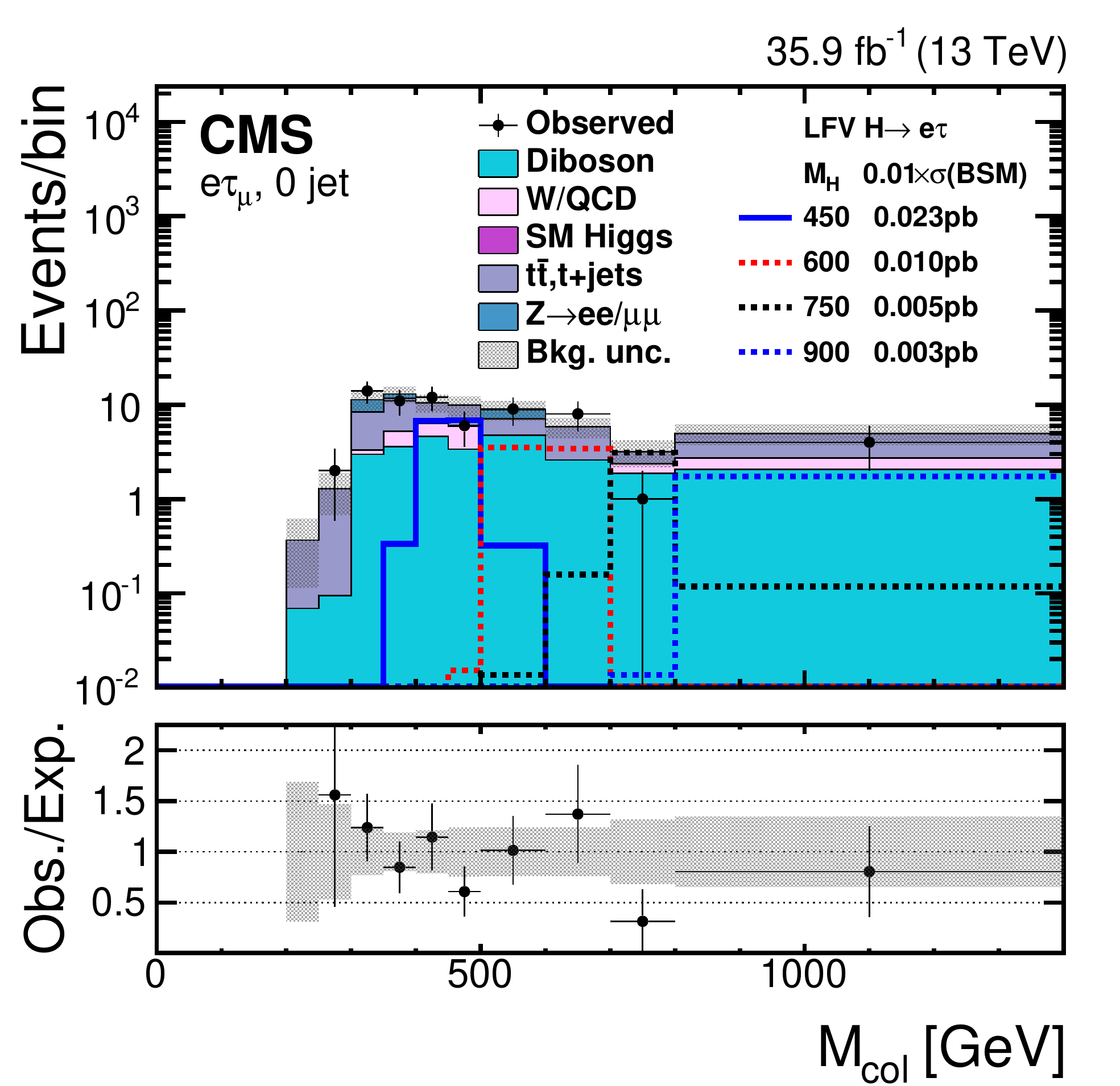}
     \includegraphics[width=0.49\textwidth]{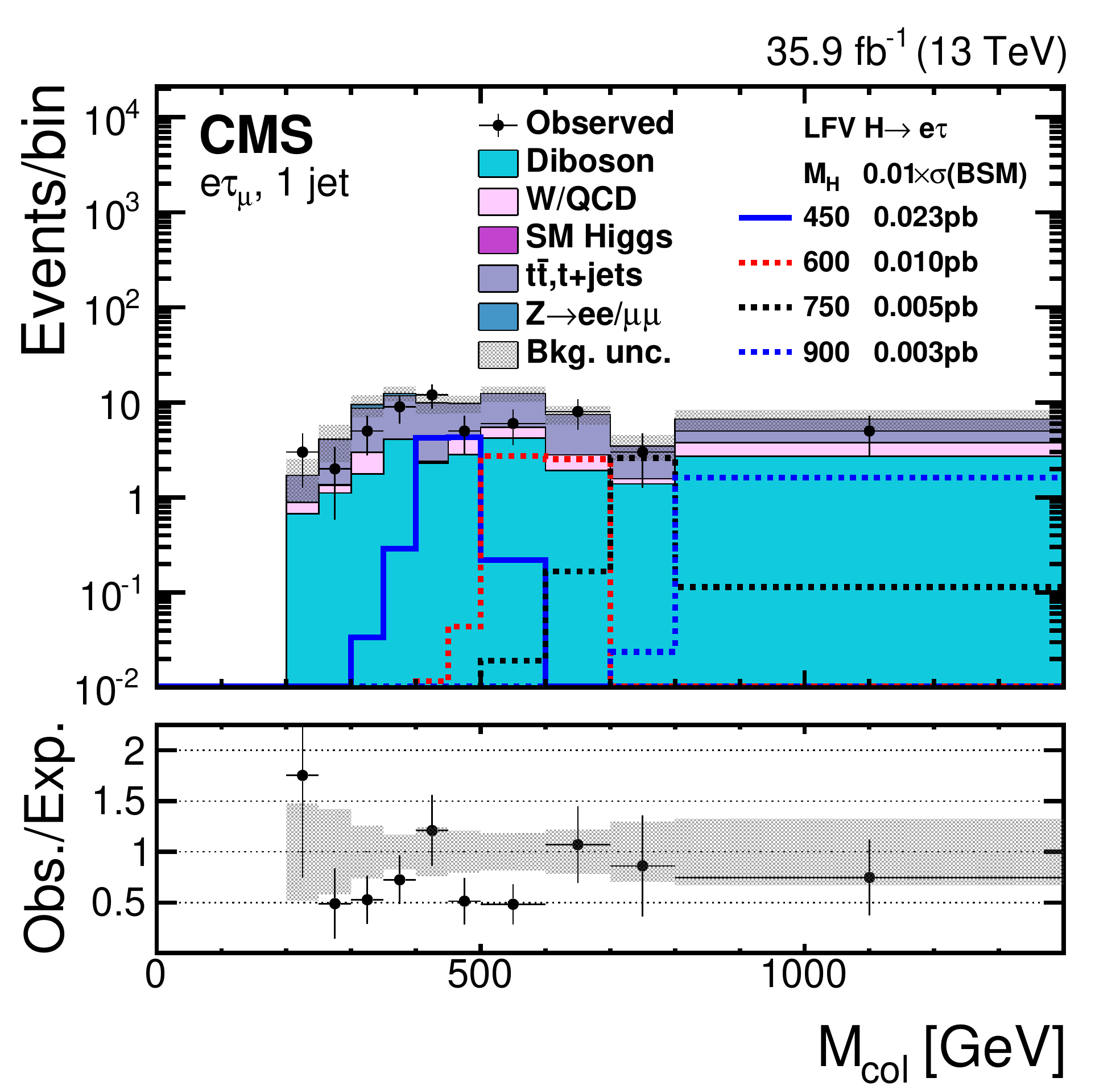}
     \caption{The \mcol distribution in the signal region, for the $\Pe\tauh$ (upper) and $\Pe\PGt_\PGm$ (lower) channels for the Higgs boson mass in the range 450--900\GeV for 0-jet (left) and 1-jet (right) categories. The uncertainty bands include both statistical and systematic uncertainties. The plotted values are number of events per bin using a variable bin size. The background is normalised to the best fit values from a binned likelihood fit, discussed in the text, to the background only hypothesis. For depicting the signals a branching fraction of 1\% and BSM cross sections from Ref.~\cite{YR4} are assumed.}
     \label{fig:McolSignalRegionETau2}
\end{figure}

\begin{figure}[htbp]
  \centering
       \includegraphics[width=0.49\textwidth]{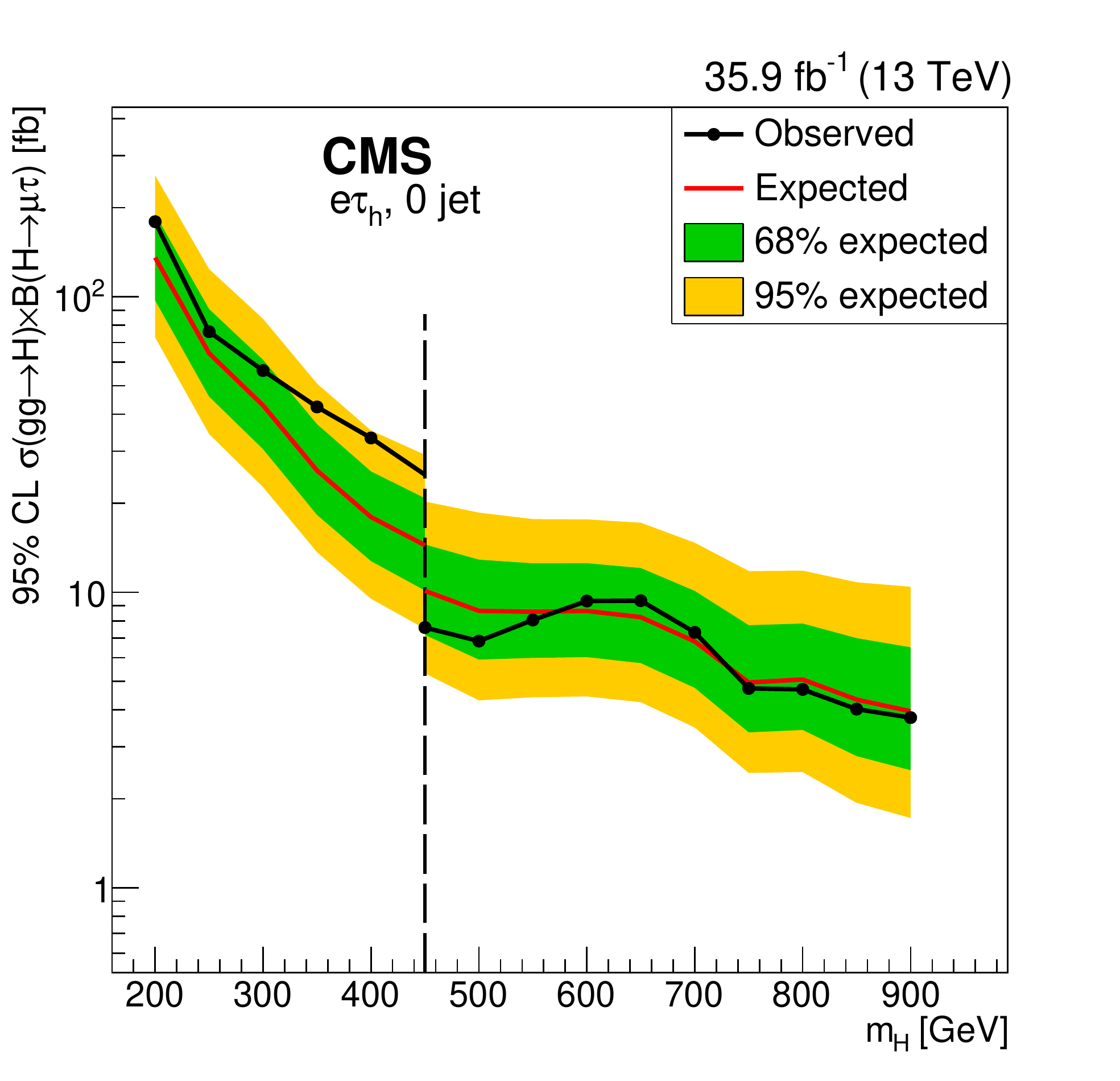}
     \includegraphics[width=0.49\textwidth]{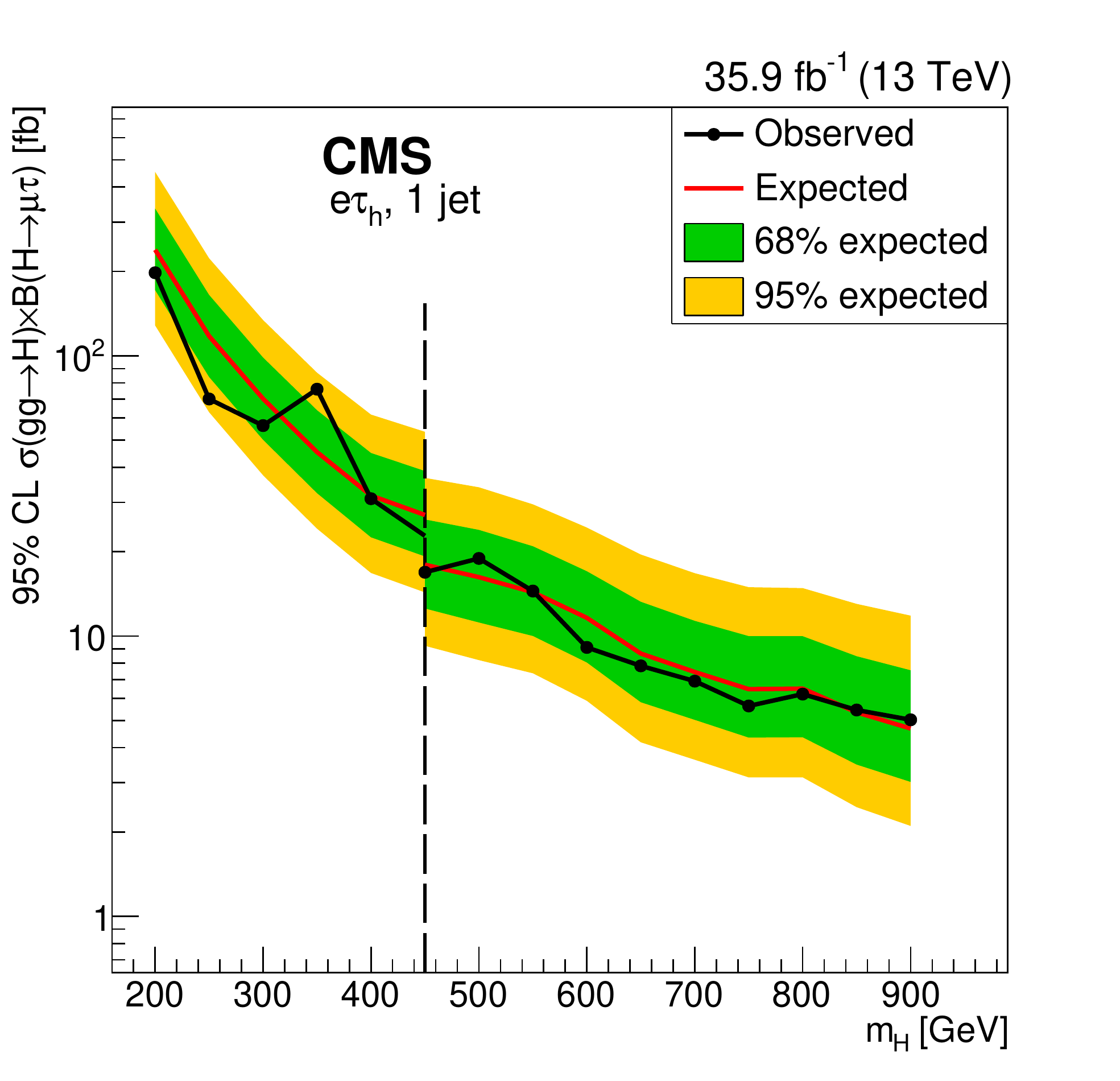}\\
     \includegraphics[width=0.49\textwidth]{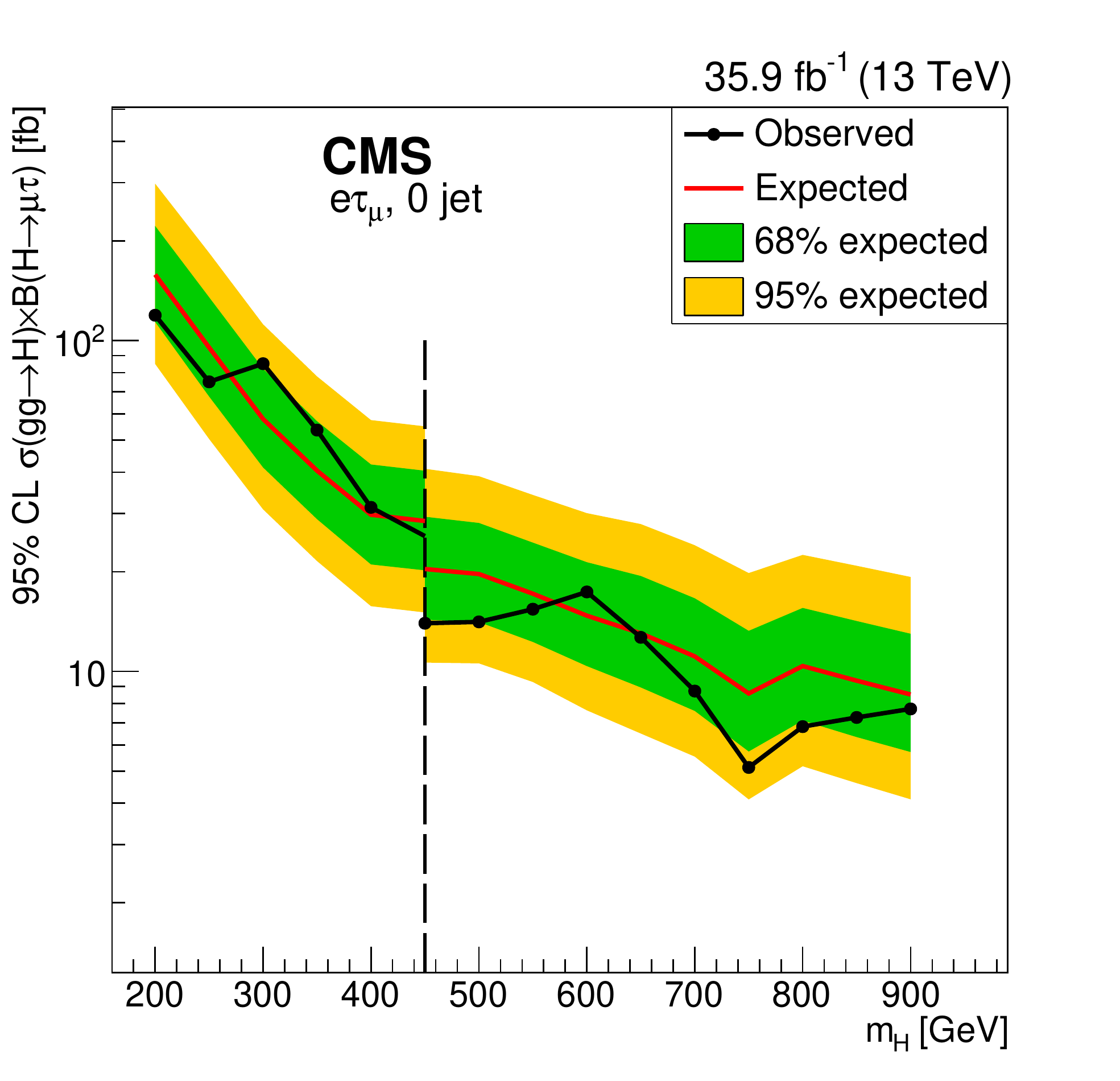}
     \includegraphics[width=0.49\textwidth]{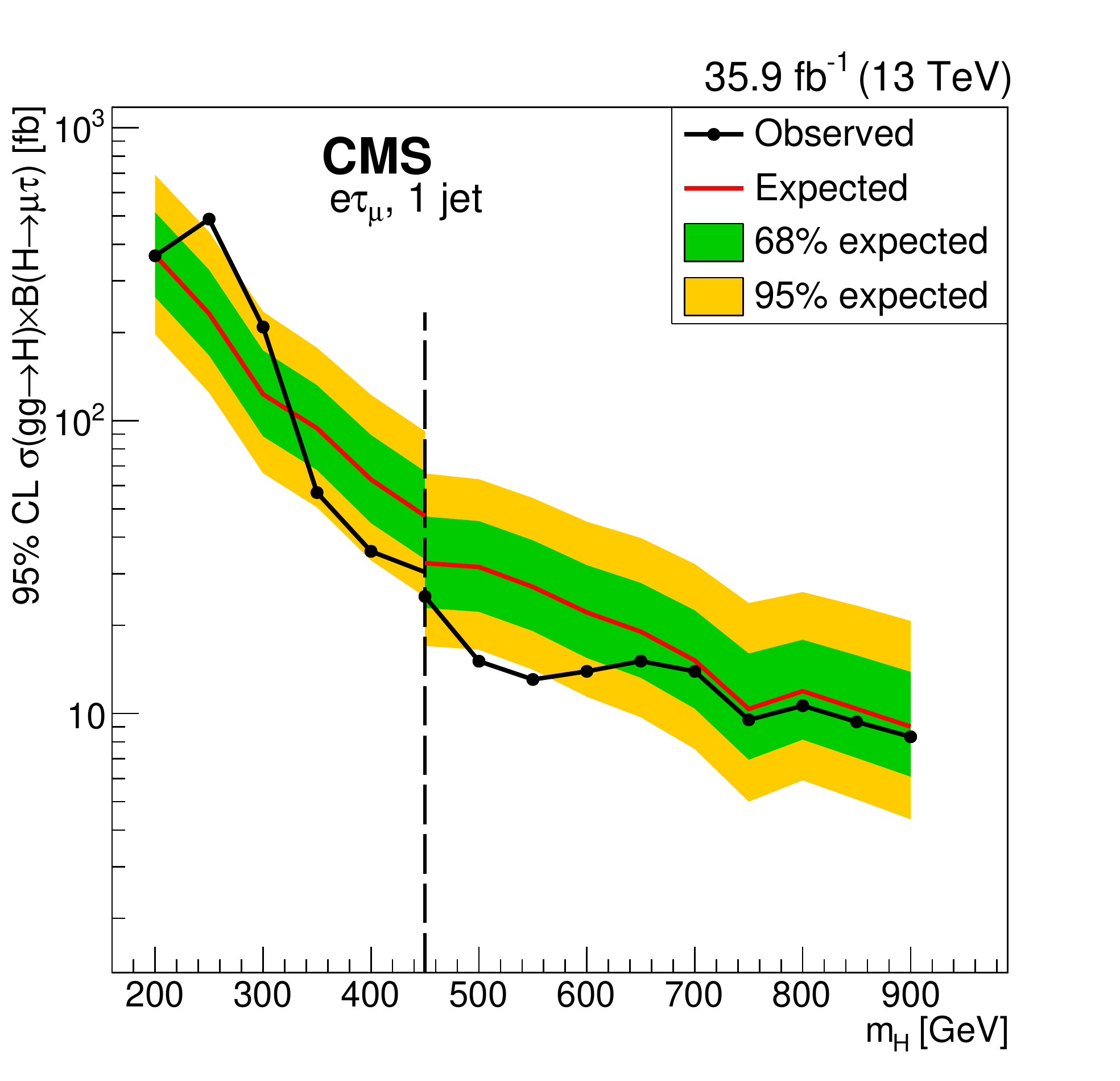}
     \caption{The observed and median expected 95\% \CL upper limits on $\sigma(\Pg\Pg\to \PH)\mathcal{B}(\PH\to\Pe\PGt)$, for the $\Pe\tauh$ (upper) and $\Pe\PGt_\PGm$ (lower) channels, for 0-jet (left) and 1-jet (right) categories. The dashed line shows the transition between the two investigated mass ranges.}
     \label{fig:masslimitsETAUcategories}
\end{figure}

\begin{figure}[htbp]
     \centering
     \includegraphics[width=0.49\textwidth]{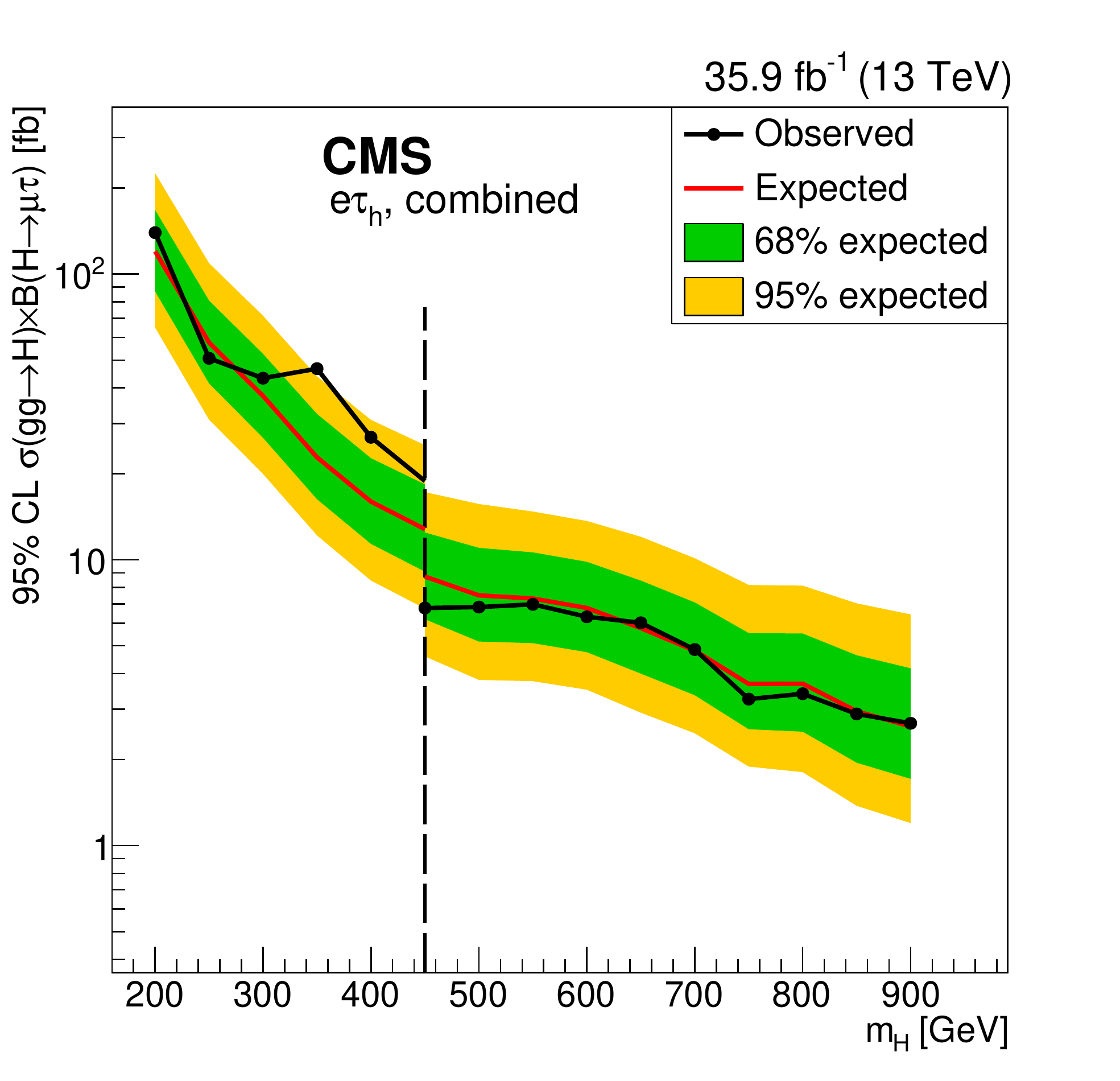}
     \includegraphics[width=0.49\textwidth]{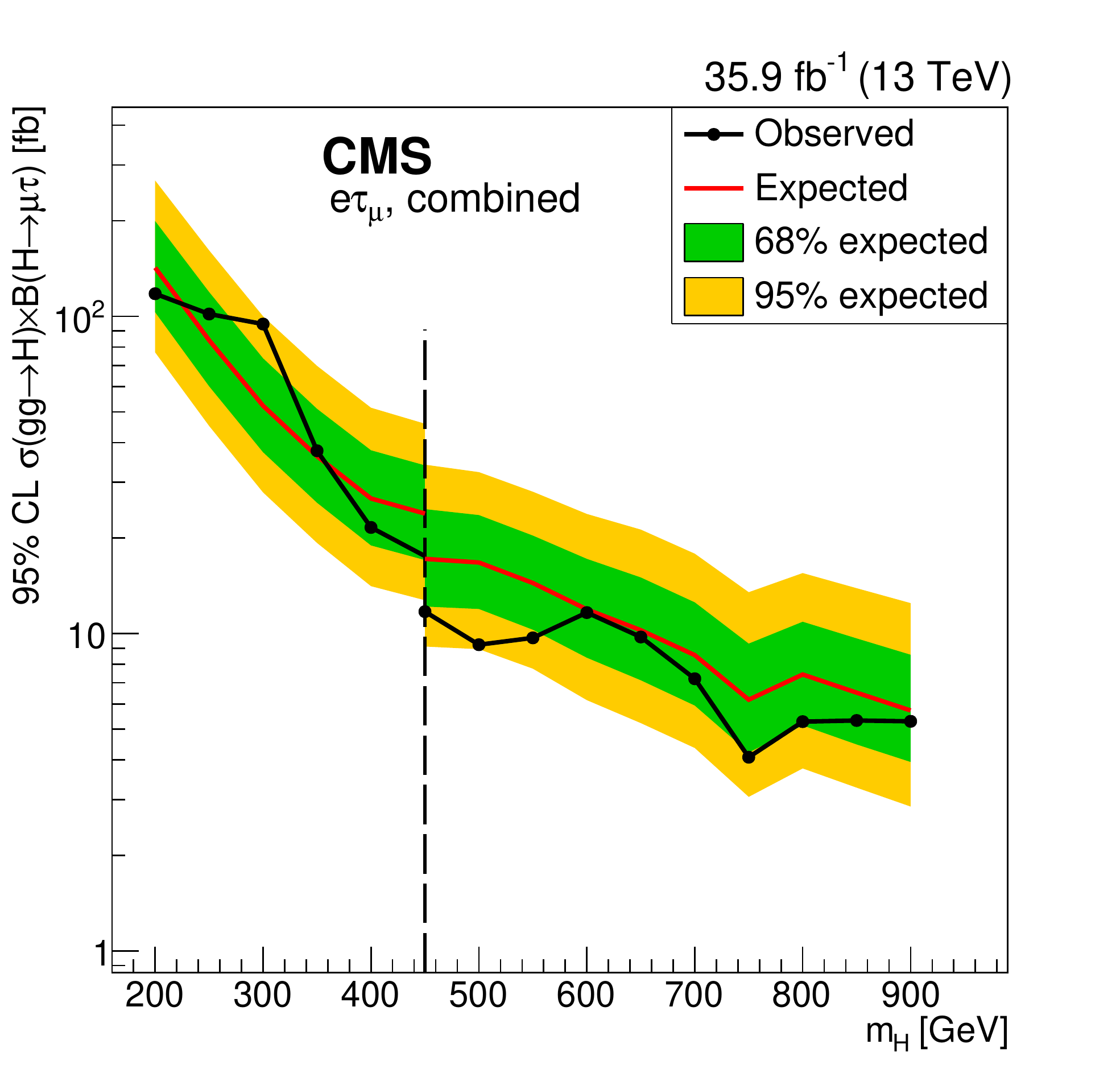}\\
     \includegraphics[width=0.8\textwidth ]{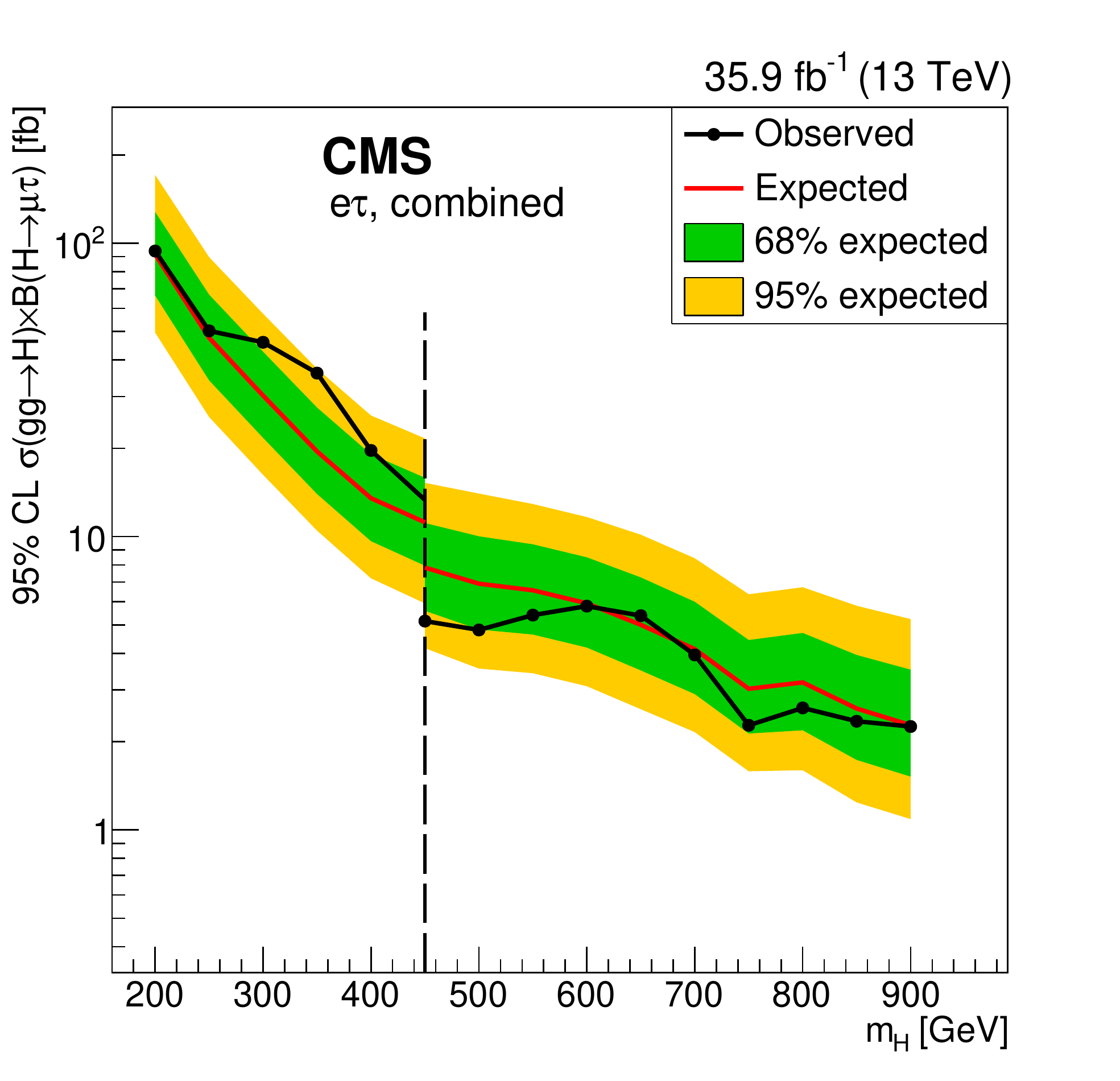}
     \caption{The combined observed and median expected 95\% \CL upper limits on $\sigma(\Pg\Pg\to \PH)\mathcal{B}(\PH\to\Pe\PGt)$, for $\Pe\tauh$ (upper left) and $\Pe\PGt_\PGm$ (upper right) channels, and their combination $\Pe\PGt$ (lower).  The dashed line shows the transition between the two investigated mass ranges.}

     \label{fig:masslimitsETAUcombined}
\end{figure}

The distributions of the collinear mass $\mcol$ compared to the signal and background contributions in the $\Hehad$ and $\Hemu$ channels, in each category, are shown in Figs.~\ref{fig:McolSignalRegionETau} and~\ref{fig:McolSignalRegionETau2}. No excess over the background expectation is observed.
The observed and median expected 95\% \CL upper limits on $\sigma(\Pg\Pg\to \PH)\mathcal{B}(\PH \to \Pe \PGt )$ range from 94.1 (91.6)\unit{fb} to 2.3 (2.3)\unit{fb}, and are given for each category in Table~\ref{tab:ETAULIMITSTABLE}. The limits are also summarized graphically in Fig.~\ref{fig:masslimitsETAUcategories} for the individual categories, and in Fig.~\ref{fig:masslimitsETAUcombined} for the combination of both two $\PGt$ decay channels.


\begin{table}
\topcaption{The observed and median expected 95\% \CL upper limits on $\sigma(\Pg\Pg\to \PH)\mathcal{B}(\PH\to\Pe\PGt)$.}\label{tab:ETAULIMITSTABLE}

\centering
\begin{tabular}{cccccccccccc}
\multicolumn{12}{c}{Observed 95\% \CL upper limit on $\sigma(\Pg\Pg\to \PH)\mathcal{B}(\PH\to\Pe\PGt)$ (fb) } \\
\hline
& \multicolumn{3}{c}{$\Pe\PGt_\PGm$} && \multicolumn{3}{c}{$\Pe\tauh$} &&\multicolumn{3}{c}{$\Pe\PGt$}  \\
$m_{\PH}$  (\GeVns) & 0 jet & 1 jet  & comb && 0 jet & 1 jet  & comb && 0 jet & 1 jet  & comb \\
\hline
200 & 119.2 & 365.3 & 117.8 && 179.4 & 197.8 & 139.6 && 103.2 & 180.1 & 94.1\\
300 & 85.1 & 208.7 & 94.5 && 56.4 & 56.4 & 43.2 & &50.6 & 65.4 & 46.0\\
450 & 14.0 & 25.1 & 11.7 && 7.6 & 16.9 & 6.8 && 5.9 & 13.2 & 5.2\\
600 & 17.4 & 13.9 & 11.7 && 9.3 & 9.1 & 6.3 && 8.8 & 6.9 & 5.8\\
750 & 5.1 & 9.5 & 4.1 && 4.7 & 5.6 & 3.3 && 2.9 & 4.5 & 2.3\\
900 & 7.7 & 8.3 & 5.3 && 3.8 & 5.0 & 2.7 && 3.1 & 4.0 & 2.3\\
\end{tabular}
\\[\cmsTabSkip]
\centering
\begin{tabular}{cccccccccccc}
\multicolumn{12}{c}{Median expected 95\% \CL upper limit on $\sigma(\Pg\Pg\to \PH)\mathcal{B}(\PH\to\Pe\PGt)$ (fb) } \\
\hline
& \multicolumn{3}{c}{$\Pe\PGt_\PGm$} && \multicolumn{3}{c}{$\Pe\tauh$} &&\multicolumn{3}{c}{$\Pe\PGt$}  \\
$m_{\PH}$ (\GeVns) & 0 jet & 1 jet  & comb && 0 jet & 1 jet  & comb && 0 jet & 1 jet  & comb \\
\hline
200 & 158.2 & 366.6 & 142.3 && 135.7 & 238.9 & 120.1  &&  102.9 & 200.5 & 91.6\\
300 & 57.9 & 123.0 & 52.3 && 42.9 & 70.3 & 37.5  &&  34.5 & 62.0 & 30.2\\
450 & 20.4 & 32.6 & 17.2 && 10.1 & 18.0 & 8.7  &&  9.0 & 15.4 & 7.8\\
600 & 14.7 & 22.1 & 11.9 && 8.6 & 11.6 & 6.8  && 7.5 & 9.9 & 5.9\\
750 & 8.6 & 10.5 & 6.2 && 4.9 & 6.5 & 3.7  &&  4.1 & 5.3 & 3.0\\
900 & 8.5 & 9.0 & 5.7 && 4.0 & 4.7 & 2.6  &&  3.3 & 4.0 & 2.3\\
\end{tabular}

\end{table}

\section{Summary}\label{sec:summary}

The first direct search for lepton flavour violating decays of a neutral non-standard-model Higgs boson (H) in the $\PGm\PGt$ and $\Pe\PGt$ channels is presented in this paper. The analyzed data set corresponds to an integrated luminosity of 35.9\fbinv of proton-proton collision data recorded at $\sqrt{s}=13$\TeV.
The results are extracted from a fit to the collinear mass  distributions. No evidence is found for lepton flavour violating decays of $\PH$ in the investigated mass range. The observed (expected) upper limits at 95\% confidence level on the product of production cross section with branching fraction, for $\PH$ mass in the range 200--900\GeV, decaying to $\PGm\PGt$ and  $\Pe\PGt$ vary from 51.9 (57.4)\unit{fb} to 1.6 (2.1)\unit{fb} and from 94.1 (91.6)\unit{fb} to 2.3 (2.3)\unit{fb}, respectively.

\begin{acknowledgments}
We congratulate our colleagues in the CERN accelerator departments for the excellent performance of the LHC and thank the technical and administrative staffs at CERN and at other CMS institutes for their contributions to the success of the CMS effort. In addition, we gratefully acknowledge the computing centres and personnel of the Worldwide LHC Computing Grid for delivering so effectively the computing infrastructure essential to our analyses. Finally, we acknowledge the enduring support for the construction and operation of the LHC and the CMS detector provided by the following funding agencies: BMBWF and FWF (Austria); FNRS and FWO (Belgium); CNPq, CAPES, FAPERJ, FAPERGS, and FAPESP (Brazil); MES (Bulgaria); CERN; CAS, MoST, and NSFC (China); COLCIENCIAS (Colombia); MSES and CSF (Croatia); RPF (Cyprus); SENESCYT (Ecuador); MoER, ERC IUT, PUT and ERDF (Estonia); Academy of Finland, MEC, and HIP (Finland); CEA and CNRS/IN2P3 (France); BMBF, DFG, and HGF (Germany); GSRT (Greece); NKFIA (Hungary); DAE and DST (India); IPM (Iran); SFI (Ireland); INFN (Italy); MSIP and NRF (Republic of Korea); MES (Latvia); LAS (Lithuania); MOE and UM (Malaysia); BUAP, CINVESTAV, CONACYT, LNS, SEP, and UASLP-FAI (Mexico); MOS (Montenegro); MBIE (New Zealand); PAEC (Pakistan); MSHE and NSC (Poland); FCT (Portugal); JINR (Dubna); MON, RosAtom, RAS, RFBR, and NRC KI (Russia); MESTD (Serbia); SEIDI, CPAN, PCTI, and FEDER (Spain); MOSTR (Sri Lanka); Swiss Funding Agencies (Switzerland); MST (Taipei); ThEPCenter, IPST, STAR, and NSTDA (Thailand); TUBITAK and TAEK (Turkey); NASU (Ukraine); STFC (United Kingdom); DOE and NSF (USA).

\hyphenation{Rachada-pisek} Individuals have received support from the Marie-Curie programme and the European Research Council and Horizon 2020 Grant, contract Nos.\ 675440, 752730, and 765710 (European Union); the Leventis Foundation; the A.P.\ Sloan Foundation; the Alexander von Humboldt Foundation; the Belgian Federal Science Policy Office; the Fonds pour la Formation \`a la Recherche dans l'Industrie et dans l'Agriculture (FRIA-Belgium); the Agentschap voor Innovatie door Wetenschap en Technologie (IWT-Belgium); the F.R.S.-FNRS and FWO (Belgium) under the ``Excellence of Science -- EOS" -- be.h project n.\ 30820817; the Beijing Municipal Science \& Technology Commission, No. Z181100004218003; the Ministry of Education, Youth and Sports (MEYS) of the Czech Republic; the Deutsche Forschungsgemeinschaft (DFG) under Germany’s Excellence Strategy -- EXC 2121 ``Quantum Universe" -- 390833306; the Lend\"ulet (``Momentum") Programme and the J\'anos Bolyai Research Scholarship of the Hungarian Academy of Sciences, the New National Excellence Program \'UNKP, the NKFIA research grants 123842, 123959, 124845, 124850, 125105, 128713, 128786, and 129058 (Hungary); the Council of Science and Industrial Research, India; the HOMING PLUS programme of the Foundation for Polish Science, cofinanced from European Union, Regional Development Fund, the Mobility Plus programme of the Ministry of Science and Higher Education, the National Science Center (Poland), contracts Harmonia 2014/14/M/ST2/00428, Opus 2014/13/B/ST2/02543, 2014/15/B/ST2/03998, and 2015/19/B/ST2/02861, Sonata-bis 2012/07/E/ST2/01406; the National Priorities Research Program by Qatar National Research Fund; the Ministry of Science and Education, grant no. 3.2989.2017 (Russia); the Programa Estatal de Fomento de la Investigaci{\'o}n Cient{\'i}fica y T{\'e}cnica de Excelencia Mar\'{\i}a de Maeztu, grant MDM-2015-0509 and the Programa Severo Ochoa del Principado de Asturias; the Thalis and Aristeia programmes cofinanced by EU-ESF and the Greek NSRF; the Rachadapisek Sompot Fund for Postdoctoral Fellowship, Chulalongkorn University and the Chulalongkorn Academic into Its 2nd Century Project Advancement Project (Thailand); the Nvidia Corporation; the Welch Foundation, contract C-1845; and the Weston Havens Foundation (USA).
\end{acknowledgments}

\bibliography{auto_generated}
\cleardoublepage \appendix\section{The CMS Collaboration \label{app:collab}}\begin{sloppypar}\hyphenpenalty=5000\widowpenalty=500\clubpenalty=5000\vskip\cmsinstskip
\textbf{Yerevan Physics Institute, Yerevan, Armenia}\\*[0pt]
A.M.~Sirunyan$^{\textrm{\dag}}$, A.~Tumasyan
\vskip\cmsinstskip
\textbf{Institut f\"{u}r Hochenergiephysik, Wien, Austria}\\*[0pt]
W.~Adam, F.~Ambrogi, T.~Bergauer, J.~Brandstetter, M.~Dragicevic, J.~Er\"{o}, A.~Escalante~Del~Valle, M.~Flechl, R.~Fr\"{u}hwirth\cmsAuthorMark{1}, M.~Jeitler\cmsAuthorMark{1}, N.~Krammer, I.~Kr\"{a}tschmer, D.~Liko, T.~Madlener, I.~Mikulec, N.~Rad, J.~Schieck\cmsAuthorMark{1}, R.~Sch\"{o}fbeck, M.~Spanring, D.~Spitzbart, W.~Waltenberger, C.-E.~Wulz\cmsAuthorMark{1}, M.~Zarucki
% ...
\vskip\cmsinstskip
\dag: Deceased\\
1:  Also at Vienna University of Technology, Vienna, Austria\\
% ...
\end{sloppypar}
\end{document}